\documentclass[11pt]{article}
\usepackage{epsf}
\usepackage{amssymb}
\usepackage[dvips]{graphicx}
\usepackage[dvips,usenames]{color}

\newcommand{\reseteqnum}{\setcounter{equation}{0}}

\title{
\hfill
\parbox{3cm}{\normalsize DPNU-01-05\\
{\tt  hep-lat/0105032}}\\
\vspace{0.5cm}
Domain wall fermion and chiral gauge theories \\
on the lattice with exact gauge invariance 
\author{
Yoshio Kikukawa\thanks{e-mail address:
kikukawa@eken.phys.nagoya-u.ac.jp} 
\\
\\
{\normalsize\em Department of Physics, Nagoya University 
}\\
{\normalsize\em Nagoya 464-8602, Japan}
\\
\\
\date{\normalsize May, 2001}
}
}

\begin{document}

\maketitle

\begin{abstract}
We discuss how to construct anomaly-free chiral gauge theories 
on the lattice with exact gauge invariance in the framework of 
domain wall fermion. Chiral gauge coupling is realized by 
introducing a five-dimensional gauge field which interpolates 
between two different four-dimensional gauge fields at boundaries. 
The five-dimensional dependence is compensated by a local and 
gauge-invariant counter term. The cohomology problem to obtain 
the counter term is formulated in 5+1 dimensional space, using 
the Chern-Simons current induced from the five-dimensional Wilson 
fermion. We clarify the connection to the invariant construction 
based on the Ginsparg-Wilson relation using overlap Dirac operator.
Formula for the measure and the effective action of Weyl fermions 
are obtained in terms of five-dimensional lattice quantities.
\end{abstract}
%\pacs{
%11.15.Ha,  \\ 12.38.Gc.
%}

\newpage
\section{Introduction}
\label{sec:intro}
\reseteqnum

Through the Ginsparg-Wilson relation \cite{ginsparg-wilson-rel}
and the exact chiral symmetry based on it \cite{exact-chiral-symmetry}, 
Weyl fermion can naturally be introduced on the lattice.
The chiral constraint imposed on the Weyl fermion is
gauge-field dependent and by introducing the basis of Weyl 
fermion, the path integral can be set 
up \cite{abelian-chiral-gauge-theory,nonabelian-chiral-gauge-theory}.
In fact, it has been shown by L\"uscher that 
the functional measure of the Weyl fermion can be constructed
in anomaly-free abelian chiral gauge theories so that it satisfies
the requirements of the smoothness, the locality and the gauge invariance
\cite{abelian-chiral-gauge-theory,topology-on-the-lattice,
effective-action-abelian-chiral,noncomutative-brs-cohomology}. The
similar construction has also been argued for  generic non-abelian
chiral gauge theories, where to treat the exact cancellation of
gauge anomaly,  a local cohomology problem in $4+2$-dimensions is 
formulated \cite{nonabelian-chiral-gauge-theory}.\footnote{
The author refers the reader to \cite{lat99-luscher,lecture-luscher-00}
for recent review of this approach. 
In this approach, 
the exact cancellation of gauge anomalies 
in non-abelian chiral gauge theories has been shown
in all orders of the expansion in lattice perturbation 
theory \cite{cohomology-non-abelian,perturbation-theory}. 
For SU(2) doublet,
it has been shown 
that Witten's global anomaly is 
reproduced \cite{oliver-isabel}. 
For 
SU(2)$_L$$\times$ U(1)$_Y$ electroweak theory, the local cohomology 
problem in $4+2$-dimensions has be solved in infinite volume lattice
and
the exact cancellation of gauge anomalies, including the mixed type,
has been shown non-perturbatively \cite{nakayama-kikukawa}.
}

This construction is generic and applies to any local lattice 
fermion theory with the Dirac operator satisfying the 
Ginsparg-Wilson relation.
In the case using the overlap Dirac operator 
\cite{overlap-D,overlap-D-GW-relation,overlap-D-odd-dim},
the path integral formalism for the Weyl fermion reproduces
the overlap formula for the 
chiral determinant by Narayanan and 
Neuberger \cite{overlap}.\footnote{
The author refers the reader to \cite{recent-review-neuberger}
for recent review of the overlap formalism.
%The author refers the reader to \cite{recent-review-neuberger}
%for recent review of the overlap formalism.
In the overlap formalism, 
reflecting chiral anomaly,
the phase factor of the chiral determinant is not fixed in general
and any reasonable choice of the phase factor should lead to 
the gauge anomaly for single Weyl fermion.
The Wigner-Brillouin phase convention has been adopted for 
perturbative studies \cite{overlap-perturbative}
and has also been tested numerically in a non-perturbative
formulation of chiral gauge 
theories \cite{overlap-nonperturbative,translational-anomaly-and-overlap}. 
Geometrical treatment of the gauge anomaly in the overlap 
formalism has been discussed in detail 
in abelian theories \cite{geometrical-aspect} and
non-abelian theories \cite{adams-global-obstructions}.
The SU(2) global anomaly has been examined 
in \cite{neuberger-su2}.
An adiabatic phase choice has been proposed 
in \cite{adiabatic-phase-chioce} and used 
in the construction of non-compact abelian chiral 
gauge theories. 
The overlap formalism in odd dimensions has been considered in
\cite{narayanan-nishimura,overlap-D-odd-dim,GW-odd-dim}.
}
The above invariant construction of the functional measure 
provides the method to fix the phase factor of the chiral
determinant in the overlap formalism
in a gauge-invariant manner.

It was also suggested
\cite{nonabelian-chiral-gauge-theory,aoyama-kikukawa} 
that there is a close relation between 
the interpolation procedure in L\"uscher's construction and
the five-dimensional setup in Kaplan's domain wall fermion 
\cite{domain-wall-fermion}.
The purpose of this paper is to 
pursue this close connection and to show how to 
construct four-dimensional lattice chiral gauge theories  
with exact gauge invariance from the five-
dimensional lattice 
framework of domain wall fermion.

For this purpose, we adopt the 
domain wall fermion in the vector-like formalism by Shamir
\cite{boundary-fermion}. But two different four-dimensional gauge fields
are introduced at the boundaries  and they are interpolated by a
five-dimensional gauge field. This inevitably causes the five-dimensional
dependence of the partition function of the domain wall fermion.
To take account of 
this five-dimensional dependence, 
we formulate an integrability condition.
It turns out that the
dependence is governed by the lattice Chern-Simons term 
induced from the five-dimensional Wilson-Dirac 
fermion (with a negative mass) \cite{3dim-CS}.

In order to compensate the five-dimensional dependence,
we require a five-dimensional counter term.
The counter term should be given by
a smooth, local and gauge invariant functional of gauge field,
in order to satisfy the requirement of the smoothness, the locality and
the gauge invariance of the low energy effective action. 
We will argue that such local, gauge-invariant field 
can be obtained in anomaly-free chiral gauge theories,
through the local cohomology problem 
in $5+1$-dimensional space  formulated with the lattice Chern-Simons 
current. 
Thus the reduction from the five-dimensional lattice to
four-dimensional lattice is acheived in a local and gauge invariant 
manner.

The locality of the lattice Chern-Simons current 
is essential for the cohomological argument in $5+1$-dimensional
space and for this 
we require the so-called admissibility condition 
\cite{
locality-of-overlap-D,
topology-on-the-lattice,abelian-chiral-gauge-theory,
nonabelian-chiral-gauge-theory} extended to
five-dimensional gauge fields (cf. \cite{locality-in-dwf}).
With this condition, several properties of the Chern-Simons current
are discussed. The earlier studies of the properties of 
the Chern-Simons current 
in the context of 
domain wall fermion can be found in 
\cite{5dim-CS-in-domain-wall-fermion,
review-chiral-dwf-jansen}.

Trying to formulate four-dimensional chiral gauge theories from the
five-dimensional framework of domain wall fermion,
our approach resembles to the wave-guide 
model \cite{waveguide-kaplan,waveguide-jansen-etal} and
the formalism proposed by Creutz et al. 
\cite{surface-mode-creutz,creutz-review}.
However, our approach is different from the wave-guide 
model 
in that we are considering 
the smooth (discrete, but smooth in lattice scale) interpolation 
in the fifth
direction.  The issue related to the disordered gauge degrees of freedom 
is taken account by the five-dimensional admissibility condition, 
which assures the existence of the chiral zero modes even 
with five-dimensional gauge fields 
(cf. \cite{waveguide-no-zeromode,overlap-not-waveguide,aoki-nagai}). 
Our approach is also different from the formalism by 
Creutz et al. 
in that we are isolating the chiral zero modes at one boundary
as physical degrees of freedom, regarding the other boundary
as reference.

This paper is organized as follows. 
In section~\ref{sec:dwf-for-chiral-gauge-theories} we formulate
the domain wall fermion for chiral gauge theories with 
the interpolating five-dimensional gauge field. 
Then we derive the integrability condition 
for the partition function of the domain wall fermion
and state a sufficient condition to obtain the five-dimensional
counter term with the required properties.
In section~\ref{sec:chern-simons-current} we examine the
properties of the lattice Chern-Simons current. 
In section~\ref{sec:reconstruction-CS-term} we argue how to
reconstruct the counter term from the Chern-Simons current
and formulate the local cohomology problem in $5+1$ dimensional space. 
Section~\ref{sec:connection-dwf-and-GW} is devoted to the 
discussions on the connection to the gauge-invariant construction
based on the Ginsparg-Wilson relation.

\section{Domain wall fermion for chiral gauge theories}
\label{sec:dwf-for-chiral-gauge-theories}
\reseteqnum

%In this section, we construct domain wall fermion 
%in chiral-asymmetric manner by introducing a 
%five-dimensional gauge field which interpolates 
%two different four-dimensional gauge fields at
%boundaries. 
%In order to take account of the 
%five-dimensional dependence of the partition function,
%an integrability condition is formulated.
%Based on it, we discuss a sufficient condition to obtain a local
%counter term which can make the partition function
%independent of the interpolation.

\subsection{Interpolation  with five-dimensional gauge field}

\begin{figure}[ht]
  \begin{center}
{\unitlength 0.7mm
\begin{picture}(160,80)

\put(15,5){\vector(1,0){120}}
\put(140,3){$x_5=t a_5 $}

\linethickness{0.9mm}
\multiput(-8,0)(80,0){2}{
  \begin{picture}(50,60)
    \put(20,5){\line(0,1){45}}
    \put(50,25){\line(0,1){45}}
    \put(20,5){\line(3,2){30}}
    \put(20,50){\line(3,2){30}}
  \end{picture}
}

\put(12,0){$t=-N+1$}
\put(92,0){$t=N$}
\put(20,35){$\Psi_L(x)$}
\put(100,35){$\Psi_R(x)$}

\put(65,35){$- m_0 $}
\put(130,35){$( + m_0 )$}
\put(-10,35){$( + m_0 )$}

\end{picture}
}
    \caption{Domain wall fermion in the simpler vector-like setup}
    \label{fig:simple-vector-like-setup}
  \end{center}
\end{figure}
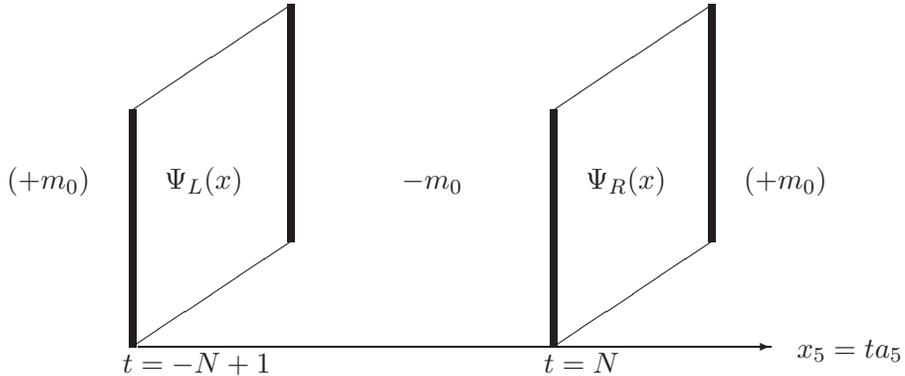

Domain wall fermion, in its simpler vector-like formulation, 
is defined by the five-dimensional Wilson-Dirac fermion 
with a negative mass in a finite extent fifth 
dimension. (See Figure~\ref{fig:simple-vector-like-setup}.)
The four-dimensional lattice spacing $a$ and the five-dimensional one
$a_5$ are both set to unity.
The fifth coordinate is denoted by $t$ and takes integer values in the
interval, $t \in [-N+1,N]$.
In four dimensions, the lattice is assumed to have a finite 
volume $L^4$ and the periodic boundary condition is assumed for
both fermion and gauge fields.
Mass term is set to the negative value $-m_0$ where $ 0 < m_0  < 2$.
\begin{equation}
S_{\rm DW} = \sum_{t=-N+1}^{N} 
             \sum_x \bar \psi(x,t) 
\left( D_{\rm 5w}-m_0 \right) \psi(x,t).
\end{equation}

\noindent
This setup is equivalent to impose the Dirichlet boundary condition 
at the boundaries in the fifth dimension as 
\begin{equation}
\label{eq:Dirichlet-bc-in-5dim}
\left.  \psi_R (x,t) \right\vert_{t=-N}  = 0, \quad
\left.  \psi_L (x,t) \right\vert_{t=N+1}  = 0.
\end{equation}

\begin{figure}[ht]
  \begin{center}
{\unitlength 1mm
\begin{picture}(160,40)
\put(20,0){\line(1,0){85}}
\put(60,-1.5){\line(0,1){3}}
\put(59.5,-5){0}

%% lattice 
\multiput(20,0)(5,0){18}{\circle*{1}}

\put(20,0){\line(0,1){30}}
\put(10,-5){$-N+1$} 
\put(105,0){\line(0,1){30}}
\put(105,-5){$N$}

% interpolation
{\linethickness{0.5mm}

\put(20,5){\line(1,0){30}}

\put(70,25){\line(1,0){35}}
\qbezier(50,5)(55,5)(60,15)
\qbezier(60,15)(65,25)(70,25)

}

\multiput(47.5,0)(0,5){6}{\line(0,1){3}}
\put(44,-5){-$\Delta$}
\multiput(72.5,0)(0,5){6}{\line(0,1){3}}
\put(72.5,-5){$\Delta$}

\put(22,8){$U^0_k(x)$}
\put(93,28){$U^1_k(x)$}
\put(50,20){$U_\mu(x,t)$}

\end{picture}
}
    \caption{Interpolating five-dimensional gauge field on the lattice}
    \label{fig:interpolation-in-fifth-dimension}
  \end{center}
\end{figure}
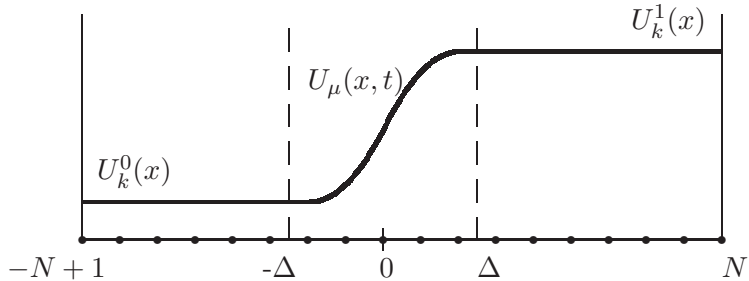

In order to introduce chiral-asymmetric gauge interaction for
the chiral zero modes at the two boundaries $t=-N+1$ and $N$, 
the gauge field is assumed to be five-dimensional, 
\begin{equation}
U_\mu(z)= \left\{ U_k(x,t) , U_5(x,t) \right\}, \quad z=(x,t)
\end{equation}
where $\mu=1,\cdots,5$ and $k=1,\cdots,4$. 
It is regarded to be interpolating a four-dimensional gauge field
at $t=-N+1$, say $U_k^0(x)$, to another four-dimensional gauge field at
$t=N$, say $U_k^1(x)$.
We assume that outside the finite interpolation region 
$t \in [-\Delta,\Delta] (\Delta < N)$  the gauge field does not depend 
on $t$ and $U_5(x,t)= 1$ 
(Figure~\ref{fig:interpolation-in-fifth-dimension}).
$\Delta$ should be chosen large enough in order 
to make sure that the interpolation is smooth enough.
The precise condition for this will be discussed below.

The gauge fields at the boundaries, $U_k^0(x)$ and $U_k^1(x)$, 
are chosen so that 
their field strengths are small enough and satisfy
the following bound 
\begin{equation}
\label{eq:smooth-gauge-field-at-boundaries}
\left\Arrowvert  1-P_{k l}(x) \right\Arrowvert 
< \epsilon, 
\qquad 
%1- 30 \epsilon > |1-m_0|^2 \quad (0 < m_0 < 2),
\epsilon < \frac{1}{30}\left\{1-|1-m_0|^2\right\},
\end{equation}
where $P_{k l}(x)$ is the four-dimensional plaquette variable.
This is the so-called admissibility 
condition \cite{locality-of-overlap-D,
topology-on-the-lattice,abelian-chiral-gauge-theory,
nonabelian-chiral-gauge-theory}.
It assures that the Hamiltonian defined through 
the transfer matrix of the five-dimensional Wilson-Dirac fermion,
$H= - \ln T$, has a finite gap and the overlap Dirac operator 
(the effective four-dimensional Dirac operator of
the boundary chiral modes) 
is local within the exponentially suppressed 
tail \cite{locality-of-overlap-D,bound-neuberger,locality-in-dwf}.
This makes the limit that the size of the fifth dimension
goes to infinity, $N\rightarrow \infty$, well-defined.

Furthermore, we also assume that the whole five-dimensional gauge field
$U_\mu(x,t)$, interpolating between $U_k^0(x)$ and $U_k^1(x)$, 
is smooth enough and satisfies the five-dimensional bound on 
the field strength as follows:
\begin{equation}
\label{eq:smooth-gauge-field-all-t}
\left\Arrowvert  1-P_{\mu\nu}(z) \right\Arrowvert 
< \epsilon^\prime, 
\qquad 
%1- 30 \epsilon > |1-m_0|^2 \quad (0 < m_0 < 2),
\epsilon^\prime < \frac{1}{50}\left\{1-|1-m_0|^2\right\}.
\end{equation}
This condition, as we will see below, assures that
the Chern-Simons current induced from the five-dimensional
Wilson-Dirac fermion is a local functional of the gauge field.
The locality of the Chern-Simons current allows us the 
cohomological analyis of the gauge anomaly in the context of 
domain wall fermion.
Since the difference of the four-dimensional
gauge fields at the boundaries is estimated as
\begin{equation}
\left\Arrowvert U_k^0(z)-U_k^1(z) \right\Arrowvert
\simeq 2 \Delta \left\Arrowvert  1-P_{5k}(z) \right\Arrowvert ,
\end{equation}
with the five dimensional bound on the field strength, 
we can make the interpolation smooth enough by taking
$\Delta$ large enough.

We also note a symmetry property of the five-dimensional
Wilson-Dirac fermion. By the reflection of the gauge field
in the fifth direction, 
\begin{eqnarray}
  U_\mu(x,t) &\rightarrow& 
U_\mu^\prime(x,t)=
\left\{ \begin{array}{l}
U_k^\prime(x,t) =U_k(x,-t+1) \\
U_5^\prime(x,t)={U_5(x,-t+1)}^{-1}
\end{array} \right. \, ,
\end{eqnarray}
the five-dimensional Wilson-Dirac operator transforms as follows:
\begin{equation}
  D_{\rm 5w} \rightarrow D_{\rm 5w}^\prime
  = P \gamma_5 D_{\rm 5w}^\dagger \gamma_5 P,
\end{equation}
where $P$ is the parity transformation operator 
in the fifth dimension:
\begin{equation}
P \ : \   t \rightarrow t^\prime = - t +1.
\end{equation}
Therefore the interpolation in the reversed order implies 
the complex conjugation.

\subsection{Integrability condition for domain wall fermion}

Now we consider the partition function of the domain wall fermion,
\begin{equation}
\left.  \det \left(D_{\rm 5w}-m_0 \right) \right\vert_{\rm Dir.}
\end{equation}
and examine its dependence on the path of the interpolation.
Let us denote by $c_1$ the original choice of the path of the interpolation, 
which is represented by the five-dimensional gauge field 
$U_\mu(x,t)$. 
In order to examine the dependence on the path, we 
introduce another path, say $c_2$ and consider the difference of 
the logarithm of the partition function:
\begin{equation}
\label{eq:dwf-partition-function-difference}
\ln \left.\det\left(D_{\rm 5w}-m_0 \right) \right\vert_{\rm Dir.}^{c_2}
-
\ln \left. \det\left(D_{\rm 5w}-m_0 \right) \right\vert_{\rm Dir.}^{c_1}.
\end{equation}

\begin{figure}[ht]
  \begin{center}
{\unitlength 0.7mm
\begin{picture}(160,40)

{\linethickness{0.5mm}

\qbezier(60,5)(80,30)(100,35)
\qbezier(60,5)(80,5)(100,35)

\put(80,5){\Large $c_1$}
\put(75,30){\Large $c_2$}

\put(52,-2){$U_k^0$}
\put(102,38){$U_k^1$}

}
\end{picture}
}
    \caption{A loop in the space of gauge fields}
    \label{fig:loop-in-fifth-dimension-I}
  \end{center}
\end{figure}
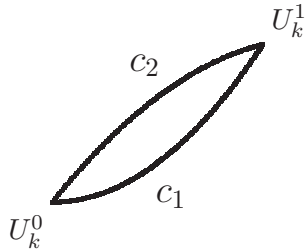

Let us assume first that 
the five-dimensional gauge field representing the path  $c_2$
can be deformed to that representing the path $c_1$,
while satisfying the constraint on the five-dimensional
plaquette variables Eq.~(\ref{eq:smooth-gauge-field-all-t}).
Whether this is always possible for any two paths 
depends on the topological structure of 
the space of the admissible gauge fields in consideration.
Since the two paths can be regareded to form a loop in the space
of the gauge fields (Fig.~\ref{fig:loop-in-fifth-dimension-I}), 
the above condition is equivalent to 
whether any loops in the space of the gauge fields
can be contractible to the point, or not.
The case with the non-contractible loops will be discussed later. 

\vspace{1em}

\noindent
We then parameterize the smooth deformation of the path
by the parameter $s \in [0,1]$  as $U_\mu^s(z)$ where
$U_\mu^{s=0}(z)=U_\mu(x)^{c_1}$ and $U_\mu^{s=1}(z)=U_\mu(z)^{c_2}$.
By differentiating the partition function 
with respect to the parameter $s$  and integrating back, 
we obtain an expression for the difference of 
the partition function, 
\begin{eqnarray}
\label{eq:dwf-partition-function-difference-integral}
&& 
\left. \ln \det \left(D_{\rm 5w}-m_0 \right) 
\right\vert_{\rm Dir.}^{c_2}
-
\left. \ln \det %{\rm Tr Ln}
\left(D_{\rm 5w}-m_0 \right) 
\right\vert_{\rm Dir.}^{c_1}
\nonumber\\
&& =
\int_0^1 ds \, \sum_x \sum_{t=-\Delta+1}^{\Delta}
\left\{ \eta_\mu^a(xt,s) \, J_\mu^a(xt,s)\vert_{\rm Dir.} \right\},
\end{eqnarray}
where $\eta_\mu^a(z,s)T^a = \partial_s U_\mu^s(z) \, {U_\mu^s(z)}^{-1}$
and $J_\mu^a(z)\vert_{\rm Dir.} $ is given by 
\begin{eqnarray}
\label{eq:dwf-gauge-current}
J_\mu^a(z)\vert_{\rm Dir.} 
&=&  
\left. {\rm Tr} 
\left( V^a_\mu(z) 
\frac{1}{D_{\rm 5w} -m_0} \right) \right\vert_{\rm Dir.} ,
\\
\label{eq:gauge-current-vertex}
V^a_\mu(z) 
&=& 
\left\{
\frac{1}{2}
\left(\gamma_\mu -1 \right) T^a U_\mu(z) \delta_{z z_1}
   \delta_{z_1 +\hat\mu, z_2}
\right. \nonumber\\
&& \qquad\qquad \left.
+
\frac{1}{2}
\left(\gamma_\mu +1 \right) U_\mu(z)^{-1} T^a \delta_{z z_2}
   \delta_{z_1 , z_2+\hat\mu}
\right\} .
\end{eqnarray}

\noindent
Since the variation of the gauge field $\eta^a_\mu(z,s)$ is 
restricted in the interpolation region 
$t \in [-\Delta+1, \Delta]$, the summation over $t$ in 
the above expression is restricted in the same finite region.

Because of this fact, as will be shown in detail in 
appendix~\ref{app:5d-wilson-D-inverse-diff},  
the difference 
Eq.~(\ref{eq:dwf-partition-function-difference-integral}) 
is well-defined and finite in the limit $N\rightarrow \infty$, as 
long as the gauge fields at the boundaries, $U_k^0$ and $U_k^1$, are 
topologically trivial and do not cause any accidental zero modes 
of the low energy effective Dirac 
operator.
%\footnote{
%In the later
%section~\ref{sec:connection-dwf-and-GW},
%we will see that this constraint on the gauge field topology is not 
%necessary and the construction works for all topological sectors.
%}
In particular, 
%as will be shown in detail in appendix~\ref{app:5d-wilson-D-inverse-diff},  
the inverse five-dimensional Wilson-Dirac operator
in Eq.~(\ref{eq:dwf-gauge-current}), 
satisfying the Dirichlet boundary condition, can be replaced 
by the inverse five-dimensional Wilson-Dirac operator
defined in the infinite extent of the fifth dimension,
\begin{equation}
\label{eq:diff-inverse-5dim-D}
\left.  \frac{1}{D_{\rm 5w} -m_0} \right\vert_{\rm Dir.} 
(xt,yt^\prime)
\quad \Longrightarrow \quad
\frac{1}{D_{\rm 5w} -m_0} \, (xt,yt^\prime),
\end{equation}
where $t,t^\prime \in [-\Delta+1,\Delta]$.
This implies that the path-dependence of 
the partition function of the domain wall fermion 
actually does not depend on the specific choice of 
the Dirichlet boundary condition,
which supports the chiral zero modes at the boundaries.

\begin{figure}[ht]
  \begin{center}
{\unitlength 0.75mm
\begin{picture}(160,40)

% base line
\put(10,0){\line(1,0){140}}
% lattice 
\multiput(10,0)(5,0){29}{\circle*{1}}

\put(0,-5){$-N+1$} 
\put(44.0,-5){0}
\put(78.0,-5){$N$}
\put(112.5,-5){$2N$}
\put(148.0,-5){$3N$}

\multiput(35,0)(0,5){6}{\line(0,1){3}}
\multiput(55,0)(0,5){6}{\line(0,1){3}}

\multiput(105,0)(0,5){6}{\line(0,1){3}}
\multiput(125,0)(0,5){6}{\line(0,1){3}}

% interpolation
{\linethickness{0.5mm}

\put(10,5){\line(1,0){25}}

\qbezier(35,5)(40,5)(45,15)
\qbezier(45,15)(50,25)(55,25)

\put(55,25){\line(1,0){50}}

\qbezier(105,25)(110,25)(113,20)
\qbezier(113,20)(120,5)(125,5)

\put(125,5){\line(1,0){25}}

}

\put(10,8){$U^0_k(x)$}
\put(140,8){$U^0_k(x)$}
\put(75,28){$U^1_k(x)$}
%\put(75,10){$U_\mu(x,t)$}

\put(43,37){$c_1,c_2$}
\put(57,37){\vector(1,0){23}}
\put(40,37){\line(-1,0){30}}

\put(10,35.5){\line(0,1){3}}
\put(80,35.5){\line(0,1){3}}
\put(150,35.5){\line(0,1){3}}

\put(111,37){$-c_0$}
\put(121,37){\vector(1,0){29}}
\put(110,37){\line(-1,0){30}}

\end{picture}
}
\vspace{1em}
    \caption{Five-dimensional gauge field representing the loop}
    \label{fig:loop-in-fifth-dimension-II}
  \end{center}
\end{figure}
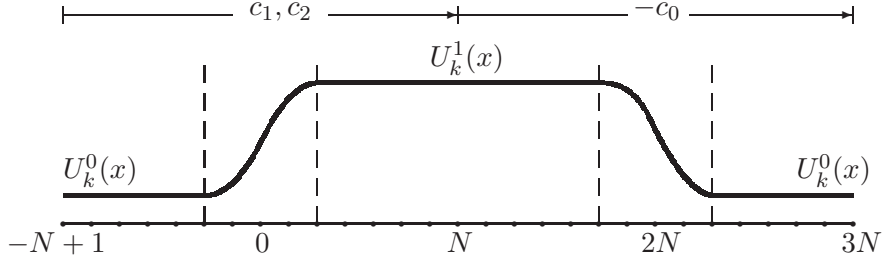

In order to make this point clear and to 
formulate an integrability condition 
for the domain wall fermion,
we introduce the five-dimensional Wilson-Dirac fermion
defined in the interval $t\in[-N+1,3N]$ in the fifth dimension
and with the anti-periodic boundary condition. 
The five-dimensional gauge field which couples to this
Wilson-Dirac fermion is assumed to form a loop in the space of 
the gauge field so that it goes along the path $c_1$ (or $c_2$) 
and comes back along a certain path $c_0$
(Fig.~\ref{fig:loop-in-fifth-dimension-II}). 
Then we can infer that, in the limit $N\rightarrow \infty$,
\begin{eqnarray}
\label{eq:dwf-integrability-condition-I}
&& 
\left. \ln \det %{\rm Tr Ln}
\left(D_{\rm 5w}-m_0 \right) \right\vert_{\rm Dir.}^{c_2}
-
\left. \ln \det %{\rm Tr Ln}
\left(D_{\rm 5w}-m_0 \right) \right\vert_{\rm Dir.}^{c_1}
\nonumber\\
&& =
\left. \ln \det %{\rm Tr Ln}
\left(D_{\rm 5w}-m_0 \right) 
\right\vert_{\rm AP}^{c_2+(-c_0)}
-
\left. \ln \det %{\rm Tr Ln}
\left(D_{\rm 5w}-m_0 \right) 
\right\vert_{\rm AP}^{c_1+(-c_0)}, \nonumber\\
&& \qquad\qquad\qquad\qquad\qquad 
\qquad\qquad\qquad\qquad\qquad 
(N \longrightarrow\infty).
\end{eqnarray}
In the r.h.s. of this identity, we may choose any paths for $c_0$.
By choosing $c_2$ and $c_1$, respectively, we can further obtain 
two identities as follows:
%\newpage
\begin{eqnarray}
{\rm Eq.(\ref{eq:dwf-integrability-condition-I})}
&=&
\left. \ln \det %{\rm Tr Ln}
\left(D_{\rm 5w}-m_0 \right) 
\right\vert_{\rm AP}^{c_2+(-c_2)}
-
\left. \ln \det %{\rm Tr Ln}
\left(D_{\rm 5w}-m_0 \right) 
\right\vert_{\rm AP}^{c_1+(-c_2)}  
\nonumber\\
&=&
\left. \ln \det %{\rm Tr Ln}
\left(D_{\rm 5w}-m_0 \right) 
\right\vert_{\rm AP}^{c_2+(-c_1)}
-
\left. \ln \det %{\rm Tr Ln}
\left(D_{\rm 5w}-m_0 \right) 
\right\vert_{\rm AP}^{c_1+(-c_1)}  .
\nonumber\\
\end{eqnarray}
By averaging these two expressions and using the
property of the five-dimensional Wilson-Dirac operator
under the reflection $t\rightarrow -t+1$, we finally obtain
\begin{eqnarray}
\label{eq:dwf-integrability-condition-II}
&& 
\left. \ln \det %{\rm Tr Ln}
\left(D_{\rm 5w}-m_0 \right) \right\vert_{\rm Dir.}^{c_2}
-
\left. \ln \det %{\rm Tr Ln}
\left(D_{\rm 5w}-m_0 \right) \right\vert_{\rm Dir.}^{c_1}
\nonumber\\
&& =
\frac{1}{2}
\left. \ln \det %{\rm Tr Ln}
\left(D_{\rm 5w}-m_0 \right) 
\right\vert_{\rm AP}^{c_2+(-c_2)}
-
\frac{1}{2}
\left. \ln \det %{\rm Tr Ln}
\left(D_{\rm 5w}-m_0 \right) 
\right\vert_{\rm AP}^{c_1+(-c_1)}
\nonumber\\
&& \quad 
+ i 
\left. {\rm Im} \ln \det %{ \, Tr \, Ln}
\left(D_{\rm 5w}-m_0 \right) \right\vert_{\rm AP}^{c_2+(-c_1)},
\quad \qquad \qquad (N \longrightarrow\infty).
\end{eqnarray}
This identity may be regarded as the integrability condition for the domain wall fermion.

Several comments about this result are in order.
First of all, in the r.h.s. of this identity, the partition functions are 
defined with the anti-periodic boundary condition and 
it is clear that there is no contribution from the low-lying chiral modes. 
They are defined well for any topological sector of $U_k^0$ and $U_k^1$.
Secondly, the first and second terms in the 
r.h.s. are turned out to be real by the reflection property
and they depend on the single paths $c_2$ and $c_1$, respectively.
These terms then can be used to subtract the bulk contribution
in the partition function of the domain wall fermion as
\begin{equation}
\left. \ln \det %{\rm Tr Ln}
\left(D_{\rm 5w}-m_0 \right) \right\vert_{\rm Dir.}^{c_i}
-  
\frac{1}{2}
\left. \ln \det %{\rm Tr Ln}
\left(D_{\rm 5w}-m_0 \right) 
\right\vert_{\rm AP}^{c_i+(-c_i)}, \ \ (i=1,2).
\end{equation}
Thirdly, an important observation 
is that the imaginary part of the r.h.s. 
is the  complex phase 
induced from the five-dimensional Wilson-Dirac fermion
with a negative mass,
which has been known to reproduce the 
Chern-Simons term in the classical continuum 
limit \cite{3dim-CS}. 
We denote this lattice Chern-Simons term 
associated with the loop $c_2+(-c_1)$ as $Q_{\rm 5w}^{c_2+(-c_1)}$:
\begin{equation}
Q_{\rm 5w}^{c_2+(-c_1)} \equiv \lim_{N\rightarrow \infty} 
{\rm Im}\ln \det %{ Tr Ln}
\left.\left(D_{\rm 5w}-m_0\right)
\right\vert_{\rm AP}^{c_2+(-c_1)}.
\end{equation}

\subsection{A sufficient condition 
for the path-independence of the domain wall fermion}

Now we discuss the requirement to obtain
the partition function of the domain wall fermion
which does not depend on the path of the interpolation 
and which  is gauge invariant.  
A sufficient condition for this can be stated as follows:

{\it 
It is possible to construct 
a local counter term which makes 
the partition function of the domain wall fermion 
independent of the path of the interpolation in a gauge-invariant manner,
if the lattice Chern-Simons term in the r.h.s. of the
integrability condition 
Eq.(\ref{eq:dwf-integrability-condition-II})
can be expressed in the form
\begin{equation}
\label{eq:CS-local-field}
  Q_{\rm 5w}^{c_2+(-c_1)}=
\lim_{N\rightarrow \infty} 
\sum_{t=-N+1}^{3N}\sum_x q_{\rm 5w}(z) , 
\end{equation}
where $q_{\rm 5w}(z)$ is a smooth, local, and gauge-invariant 
functional of the five-dimensional gauge field, 
for all possible loops in the space of gauge fields.} 

\vspace{1em}
\noindent
We first assume that this is the case 
and discuss the consequences of this condition.
In the next sections, we will discuss how to obtain the 
local and gauge-invariant functional $q_{\rm 5w}(z)$.

\subsubsection{Local counter term}
The immediate consequence of the localization property of $q_{\rm 5w}(z)$ 
is that 
$Q_{\rm 5w}^{c_2+(-c_1)}$ can be 
decomposed into two parts 
$C_{\rm 5w}^{c_2}$ and $C_{\rm 5w}^{(-c_1)}=-C_{\rm 5w}^{c_1}$, 
which are associated with the paths $c_1$ and $c_2$, respectively,
and do not depend on other paths. 
Namely,
\begin{equation}
\label{eq:CS-decomposition}
 Q_{\rm 5w}^{c_2+(-c_1)}
=C_{\rm 5w}^{c_2}  -C_{\rm 5w}^{c_1},
\end{equation}
where
\begin{eqnarray}
C_{\rm 5w}^{c_1}
&=& \lim_{N\rightarrow \infty} \sum_{t=-N+1}^{N}\sum_x  \,
q_{\rm 5w}(z) , \\ 
-C_{\rm 5w}^{c_2}
&=& \lim_{N\rightarrow \infty} \sum_{t=+N+1}^{3N}\sum_x \, q_{\rm 5w}(z) .
\end{eqnarray}

Then from the integrability condition 
Eq.~(\ref{eq:dwf-integrability-condition-II})
we can infer that in the limit $N\rightarrow \infty$
\begin{equation}
\label{eq:dwf-partition-function-path-independence}
%\lim_{N\rightarrow \infty}
\frac{ \left. \det \left(D_{\rm 5w}-m_0\right) 
\right\vert_{\rm Dir.}^{c_2} }
{ \left\vert \left. \det \left(D_{\rm 5w}-m_0\right) 
\right\vert_{\rm AP} ^{c_2+(-c_2)}
\right\vert^{\frac{1}{2}} }  \, e^{-iC_{\rm 5w}^{c_2}}
= 
%\lim_{N\rightarrow \infty}
\frac{ \left. \det \left(D_{\rm 5w}-m_0\right) \right\vert_{\rm Dir.}^{c_1} }
{ \left\vert \left. \det \left(D_{\rm 5w}-m_0\right) 
\right\vert_{\rm AP} ^{c_1+(-c_1)}
  \right\vert^{\frac{1}{2}} }
\, e^{-iC_{\rm 5w}^{c_1}} .
\end{equation}
This holds for any two paths. Therefore the 
(subtracted) partition function of the domain wall fermion
plus the local counter term does not 
depend on the path of the interpolation and is 
determined uniquely by the four-dimensional gauge fields,
$U_k^0$ and $U_k^1$. 
Thus the reduction from the five-dimensional lattice to 
the four-dimensional lattice is achieved.

We will see 
in section~\ref{sec:connection-dwf-and-GW} that
this result holds for all topological sectors,
through the more direct calculation of the partition function 
of the domain wall fermion.

\subsubsection{Gauge invariance}
Let us examine the gauge transformation property of the 
(subtracted) partition function of the domain wall fermion
with the local counter term 
under the gauge transformation for $U_k^1$,
\begin{equation}
  U_k^1(x) \rightarrow {}^g U_k^1(x) =
g(x)  U_k^1(x) g(x+\hat k)^{-1} .
\end{equation}
Since 
the partition function 
does not depend on the interpolation 
between ${}^gU^1$ and $U^0$, one may choose
any interpolation path between them.  In this case, it turns out to be 
convenient to choose the interpolation as 
follows: we first interpolate the gauge function as
\begin{equation}
  G(z) = \left\{ 
    \begin{array}{cl}
      g(x) & t \in [\Delta+1,N] \\
      g(x,t) & t \in [-\Delta+1,\Delta] \\
      1      & t \in [-N+1, -\Delta]
    \end{array}
   \right.
\end{equation}
and then apply it as a five-dimensional gauge transformation
to the %five-dimensional 
gauge field representing the original interpolation 
between $U_k^1$ and $U_k^0$.
\begin{equation}
  U_\mu(z) \rightarrow {}^GU_\mu(z)=G(z)U_\mu(z) G(z+\hat\mu)^{-1} . 
\end{equation}
In Eq.~(\ref{eq:dwf-partition-function-path-independence}), 
the partition functions of the five-dimensional Wilson-Dirac fermions
with both boundary conditions are invariant under such
five-dimensional gauge transformation. Then 
the gauge transformation is given solely by the gauge variation of the 
counter term associated with the path $c_1$.
\begin{equation}
\delta_G C_{\rm 5w}^{c_1}
 = \lim_{N\rightarrow \infty} 
\sum_{t=-N+1}^{N}\sum_x  \delta_G \, q_{\rm 5w}(z) . 
\end{equation}
Thus the question of the gauge invariance of 
the system
reduces to the question of the gauge 
invariance of the counter term associated with the path $c_1$.
When $q_{\rm 5w}(z)$ is a gauge invariant local field, 
it is gauge invariant.

\section{Properties of lattice Chern-Simons current}
\label{sec:chern-simons-current}
\reseteqnum

In this section, we discuss properties of 
the Chern-Simons current which is obtained 
from the lattice Chern-Simons term by the variation with respect to 
gauge field. We will argue that the lattice Chern-Simons current is
a smooth and local functional of the gauge field. 

\subsection{Chern-Simons current}

Let us consider 
the smooth deformation of the five-dimensional gauge field $U_\mu(z)$
representing the loop in the Chern-Simons term. Under the deformation, 
\begin{equation}
\delta U_\mu(z) \, U_\mu(z)^{-1} = T^a \eta_\mu^a (z) ,
\qquad 
{\rm Tr} \left\{ T^a T^b \right\}= - \frac{1}{2} \delta^{ab} ,
\end{equation}
the variation of the 
lattice Chern-Simons term is given by 
\begin{eqnarray}
\delta  \, 
\left. {\rm Im Tr Ln}\left(D_{\rm 5w}-m_0\right) \right\vert_{AP}
&=& 
{\rm Im} \, 
{\rm Tr} 
\left.
\left( \delta D_{\rm 5w} \frac{1}{D_{\rm 5w} -m_0} \right) 
\right\vert_{AP} 
\\
&=&
\sum_z 
\eta_\mu^a(z) \, J_\mu^a(z)
\end{eqnarray}
where
\begin{eqnarray}
J_\mu^a(z) 
=  
\left. {\rm Im} \, {\rm Tr} 
\left( V^a_\mu(z) 
\frac{1}{D_{\rm 5w} -m_0} \right) \right\vert_{AP} ,
\end{eqnarray}
\begin{eqnarray}
&&V^a_\mu(z) 
= 
\left\{
\frac{1}{2}
\left(\gamma_\mu -1 \right) T^a U_\mu(z) \delta_{z z_1}
   \delta_{z_1 +\hat\mu, z_2}
\right. \nonumber\\
&& \qquad\qquad\qquad \left.
+
\frac{1}{2}
\left(\gamma_\mu +1 \right) U_\mu(z)^{-1} T^a \delta_{z z_2}
   \delta_{z_1 , z_2+\hat\mu}
\right\} .
\end{eqnarray}
We refer the current $J_\mu^a(z)$ as Chern-Simons current.

Since the Chern-Simons current is defined by the variation of 
the Chern-Simons term which is the gauge-invariant 
functional of the five-dimensional gauge field, it is 
gauge-covariant, integrable and conserved:\footnote{
$\partial_\mu$ and $\partial_\mu^\ast$ 
denote the forward and backward difference operators, respectively,
\[
\partial_\mu f(x) = \sum_\mu \{ f(x+\hat \mu)-f(x) \}, \quad
\partial_\mu^\ast f(x) = \sum_\mu \{ f(x)-f(x-\hat \mu) \},
\]
while $D_\mu$ and $D_\mu^\ast$ denote the covariant counterparts,
\[
D_\mu f(z) = \sum_\mu \{ U_\mu(z) f(z+\hat \mu)U_\mu(z)^{-1}-f(z) \},
\quad 
D_\mu^\ast f(z) = \sum_\mu f(z)-U_\mu(z-\hat \mu)f(z-\hat \mu)
U_\mu(z-\hat \mu)^{-1}.
\]
}
\begin{equation}
\label{eq:CS-current-integrability}
\sum_z %\sum_t \sum_x  %=-\Delta+1}^\Delta
\eta_\mu^a(z) \, \delta_\zeta J_\mu^a(z)
- 
\sum_z %\sum_t \sum_x  %=-\Delta+1}^\Delta
\zeta_\mu^a(z) \, \delta_\eta J_\mu^a(z)
- 
\sum_z %\sum_t \sum_x  %=-\Delta+1}^\Delta
\left\{ [ \eta_\mu(z),\zeta_\mu(z)] \right\}^a \, J_\mu^a(z)
= 0,
\end{equation}
\begin{equation}
\left\{ D_\mu^\ast J_\mu\right\}^a (z) = 0 .
\end{equation}

\subsection{Locality of the Chern-Simons current}

Next we argue that
the Chern-Simons current is a local functional of the five-dimensional
gauge field,
as long as the constraint on the five-dimensional
plaquette variables Eq.~(\ref{eq:smooth-gauge-field-all-t}) is full-filled.
This fact follows from the following consideration.
With the bound Eq.~(\ref{eq:smooth-gauge-field-all-t}), we infer that 
the five-dimensional Wilson-Dirac operator 
%in the infinite extent of the fifth dimension 
is bounded from below 
by a positive constant 
\cite{locality-of-overlap-D,locality-in-dwf},
\begin{equation}
%\left\| %a^2 
\left(D_{\rm 5w}- m_0\right)^\dagger 
            \left(D_{\rm 5w}- m_0 %\frac{m_0}{a}
\right)
%\right\|
\ge \,  
\left\{ (1- 50 \epsilon )^{\frac{1}{2}}-|1-m_0| \right\}^2 .
\end{equation}
Given the positive lower and upper bounds for the five-dimensional
Wilson-Dirac operator, 
\begin{equation}
\tilde \alpha \le
%\left\| %a^2 
\left(D_{\rm 5w}- m_0 %\frac{m_0}{a}
                              \right)^\dagger 
            \left(D_{\rm 5w}- m_0 %\frac{m_0}{a}
                              \right)
%\right\|
\le \tilde \beta,  
\end{equation}
it follows that 
the inverse five-dimensional Wilson-Dirac operator 
in the infinite lattice decays
exponentially at large distance in the five 
dimensions \cite{locality-of-overlap-D,locality-in-dwf}:
\begin{equation}
\label{eq:exponetial-bound-on-5dim-D}
\left\| \left\{ %a^2 
D_{\rm 5w}^\dagger  
D_{\rm 5w}\right\}^{-1}(z,w) 
\right\| \le  \, 
C \, 
\exp \left\{ - \frac{\tilde \theta}{2}d_5(z,w)\right\} ,
\end{equation}
where $d_5(z,w)=|z-w|=|x-y| +|s-t|$ and
\begin{equation}
C=\frac{4 t}{\tilde \beta-\tilde \alpha} 
\left( 
\frac{1}{1-t} \frac{d_5(z,w)}{2} + \frac{t}{(1-t)^2} 
\right),
\end{equation}
\begin{equation}
t= e^{-\tilde \theta}, \qquad \cosh 
\tilde \theta = \frac{\tilde \beta + \tilde \alpha}
              {\tilde \beta - \tilde \alpha} .
\end{equation}
Similar exponential bound can be established for 
the differentiation with respect to the gauge field.

The Chern-Simons current in consideration is defined in 
the finite volume lattice. It can be expressed in terms 
of the inverse five-dimensional Wilson-Dirac operator 
in the infinite lattice as 
\begin{eqnarray}
J_\mu^a(z)
&=& 
\sum_{n \in Z^5} (-)^{n_5}
{\rm Im} \, {\rm tr} 
\left( 
\frac{1}{2}
\left(\gamma_\mu -1 \right) T^a U_\mu(z)
%V^a_\mu(z;w_1,w_2) 
\frac{1}{D_{\rm 5w} -m_0} (z+\hat \mu,z+n \cdot L )\right.
\nonumber\\
&& \qquad\qquad \left.
+\frac{1}{2}
\left(\gamma_\mu +1 \right) U_\mu(z)^{-1} T^a
%V^a_\mu(z;w_1,w_2) 
\frac{1}{D_{\rm 5w} -m_0} (z,z+\hat \mu+n \cdot L )\right),
\nonumber\\
\end{eqnarray}
where $n \cdot L = (\sum_k n_k) L + n_5 (4N)$.
Then we see that the dependence of the Chern-Simons
current on the gauge field is exponentially suppressed 
at large distance in the leading contribution with $n=0$,
while the remaining dependences on the gauge field are at most
of order ${\cal O}( \exp(-\theta \, L/2),\exp(-\theta \, 2N) )$ 
and exponentially small.  In this sense, the Chern-Simons current 
in the finite volume lattice can be regarded as a local functional 
of gauge field.

\subsection{The Chern-Simons current and gauge anomaly}

If we consider the variation of the Chern-Simons term
under the gauge transformation, we obtain
\begin{equation}
\delta  \, 
\left. {\rm Im Tr Ln}\left(D_{\rm 5w}-m_0\right) \right\vert_{AP}
= 
\sum_z %\sum_t \sum_x  %=-\Delta+1}^\Delta
\{D_\mu \omega(z)\}^a\, J_\mu^a(z)
= \sum_z \partial_\mu \left\{ \omega^a(z) J_\mu^a(z) \right\} .
\nonumber\\
\end{equation}
The flow of the Chern-Simons current should be responsible for
the gauge anomaly associated with the chiral zero mode at the
boundaries \cite{callan-harvey}. 
In fact, it has been shown by Golterman, Jansen and Kaplan 
\cite{5dim-CS-in-domain-wall-fermion}
that 
the asymptotic value of the fifth component of the Chern-Simons current
reproduces the known result of the gauge anomaly 
in the classical continuum limit:
\begin{eqnarray}
\label{eq:CS-current-gauge-anomaly}
\lim_{a \rightarrow 0} \lim_{N\rightarrow \infty}  J_5^a(x,N) 
&=& - \frac{1}{32\pi^2} \epsilon_{klmn}
{\rm Tr} \left\{ T^a F_{kl}^1(x) F_{mn}^1(x) \right\}.
\end{eqnarray}

\section{Reconstruction of the Chern-Simons term  
from the Chern-Simons current}
\label{sec:reconstruction-CS-term}
\reseteqnum

In this section, we discuss how
to obtain the local and gauge invariant field of the Chern-Simons term. 
Using the Chern-Simons current obtained in the previous section, 
we introduce a local topological field on $5+1$-dimensional space and
formulate a local cohomology problem. We will see that 
the trivial solution of the cohomology problem leads to the 
local field with the required properties.

\subsection{Contractible loops}

Let us first assume that 
a loop $l$ in the space of the gauge fields can 
be contractible to a point. 
Namely, we assume that the
five-dimensional gauge field $U_\mu(z)$ representing the loop $l$
can be deformed to the uniform gauge field 
$U_\mu(z)=U_k^0(x)$ (for all $t$),
while satisfying the constraint on the five-dimensional
plaquette variables Eq.~(\ref{eq:smooth-gauge-field-all-t}).
It depends on the topological structure of 
the space of the admissible gauge fields in consideration, 
whether all possible loops are contractible or not. The case 
with the non-contractible loops will be discussed later. 

Let us parameterize the smooth deformation of the loop $l$ 
by the parameter $s \in [0,1]$ as $U_\mu^s(z)$, where
$U_\mu^{s=0}(z)=U_k^0(x)$ and $U_\mu^{s=1}(z)=U_\mu(z)$.
The trivial interpolation (the point) at $s=0$ is denoted by $l^0$.
By differentiating the Chern-Simons term 
with respect to the parameter $s$  and then integrating back, 
we obtain an expression for the Chern-Simons term as
\begin{eqnarray}
\label{eq:reconstructed-CS}
{\rm Im Tr Ln} \left.\left(D_{\rm 5w}-m_0\right)\right\vert_{\rm AP}(l)
&=&   
\sum_z
\left\{ \int_0^1 ds \, \eta_\mu^a(z,s) \, J_\mu^a(z)\vert_{U_\mu^s} \right\},
\end{eqnarray}
where $\eta_\mu^a(z,s)T^a = \partial_s U_\mu^s(z) \, {U_\mu^s(z)}^{-1}$.
Here we took account of the fact that for the point $l^0$ 
represented by the uniform 
gauge field $U_\mu^0(x)$, the Chern-Simons term vanishes 
identically, ${\rm Im Tr Ln} 
\left.\left(D_{\rm 5w}-m_0\right)\right\vert_{\rm AP}(l^0)=0$,
because of the reflection property of the five-dimensional
Wilson-Dirac operator discussed in 
section~\ref{sec:dwf-for-chiral-gauge-theories}.

The Chern-Simons term so reconstructed is now given by 
the sum of the local field in the curly bracket,
because the Chern-Simons current is a local functional of the gauge field.
We note, however, that the local field is not gauge
invariant. In fact, under an infinitesimal gauge transformation,
$\delta U_\mu^s \left\{U_\mu^s\right\}^{-1}= -D_\mu \omega(z,s) $,
the variation of the local field is given by
\begin{eqnarray}
\label{eq:reconstructed-CS-variation}
  \delta
\left\{ \int_0^1 ds \, \eta_\mu^a(z,s) \, J_\mu^a(z)\vert_{U_\mu^s} \right\}
&=& 
-\int_0^1 ds \, \left\{D_\mu  \partial_s \omega(z,s)\right\}^a 
\, J_\mu^a(z)\vert_{U_\mu^s} . \nonumber\\
\end{eqnarray}

The clue to obtain the gauge invariant local field
is to note that the field in 
Eq.~(\ref{eq:reconstructed-CS-variation}) 
may be corrected  by the total divergence of a certain 
local current $K_\mu(z)$ without affecting
the Chern-Simons term:
\begin{equation}
\label{eq:reconstructed-CS-corrected}
{\rm Im Tr Ln} \left.\left(D_{\rm 5w}-m_0\right)\right\vert_{\rm AP}(l)
=   
\sum_z \int_0^1 ds \, 
\left\{ \eta_\mu^a(z,s) \, J_\mu^a(z)\vert_{U_\mu^s}  
- \partial_\mu^\ast K_\mu (z)\vert_{U_\mu^s,\eta_\mu} 
\right\}.
\end{equation}
The local field can be made gauge invariant if $J_\mu^a(z)$ 
would satisfy the relation
\begin{equation}
\label{eq:condition-for-gauge-invariance}
\delta 
\left\{ \eta_\mu^a(z,s) \, J_\mu^a(z)\vert_{U_\mu^s} \right\}
= 
-\left\{D_\mu  \partial_s \omega(z,s)\right\}^a 
\, J_\mu^a(z)\vert_{U_\mu^s}  
=\partial_\mu^\ast \, \delta K_\mu(z)\vert_{U_\mu^s,\eta_\mu} 
\end{equation}
with a certain local current $K_\mu(z)$,
under the infinitesimal gauge transformation
$\delta U_\mu^s \left\{U_\mu^s\right\}^{-1}= -D_\mu \omega(z,s)$.
As we can see from Eq.~(\ref{eq:condition-for-gauge-invariance}),
the question whether $J_\mu^a(z)$ would satisfy the above
relation and how to find the local current $K_\mu(z)$
defines a local cohomology problem. 

This cohomology problem can be reformulated as a local 
cohomology problem in higher dimensions \cite{nonabelian-chiral-gauge-theory}.
In the next subsection, we formulate the cohomology problem 
in five-dimensinal lattice plus one dimensional continuum space.

\subsection{Cohomology problem in $5+1$ dimensional space}

In order to reformulate the local cohomology problem 
in five-dimensinal lattice plus one dimensional continuum space,
let us introduce a gauge field on the $5+1$-dimensional space as 
\begin{equation}
\left(  U_\mu(z,s),A_6(z,s) \right), 
\end{equation}
where $A_6=T^a A_6^a$ and its gauge transformation property 
is specified as 
\begin{equation}
  A_6(z,s) \longrightarrow {}^G A_6(z,s) = G(z,s) A_6(z,s) G(z,s)^{-1}
                                     - \partial_s G(z,s) \, G(z,s)^{-1}.
\end{equation}
Accordingly, the covariant derivative in the continuum dimension can be
defined by
\begin{equation}
{\cal D}_s U_\mu(z,s) = \partial_s U_\mu(z,s)
                         + A_6(z;s) U_\mu(z,s)
                         - U_\mu(z,s) A_6(z+\hat\mu,s).
\end{equation}

We now introduce a gauge invariant  and local field 
{\it in the 5+1 dimensional space} by 
\begin{eqnarray}
\label{eq:5+1-dim-topological-field-1}
q(z,s)&\equiv& 
\left\{ {\cal D}_s U_\mu(z,s) \, U_\mu(z,s)^{-1} \right\}^a
%\eta_\mu^a(z,s) 
\left. J_\mu^a(z) \right\vert_{U_\mu^s}  \\
&=& 
\label{eq:5+1-dim-topological-field-2}
%\left\{ \partial_s U_\mu(z,s) \, U_\mu(z,s)^{-1} \right\}^a
\eta_\mu^a(z,s) 
\left. J_\mu^a(z) \right\vert_{U_\mu^s}  
- \left\{ D_\mu A_6(z,s) \right\}^a
\left. J_\mu^a(z) \right\vert_{U_\mu^s}  ,
\end{eqnarray}
where $
T^a \eta_\mu^a (z,s) = \partial_s U_\mu(z,s) \, U_\mu(z,s)^{-1}$.
This local field is topological. 
Namely, the summation of the field 
over the $5+1$-dimensional space is invariant under the 
local variation of the $5+1$-dimensional gauge field:
\begin{equation}
  \sum_z \int_0^1 ds \,\, \delta q(z,s) = 0 .
\end{equation}
In fact, denoting 
$\delta U_\mu \, U_\mu^{-1}=T^a \zeta_\mu^a$ and 
using Eq.~(\ref{eq:CS-current-integrability}), 
it follows from the second expression of the topological field 
in Eq.~(\ref{eq:5+1-dim-topological-field-2}) that 
\begin{equation}
\sum_z \int_0^1 ds \delta q(z,s) 
=\sum_z \int_0^1 ds \left( 
\partial_s \left\{
\zeta_\mu^a J_\mu^a\right\}
- \partial_\mu \, \delta \left\{ A_6^a J_\mu^a \right\}
\right) .
\end{equation}

Now let us assume that 
this topological field is cohomologically trivial, 
that is, it can be written in the form
\begin{equation}
\label{eq:5+1-dim-topological-field-trivial}
q(z,s) = \partial_\mu^\ast k_\mu(z,s) + \partial_s k_6(z,s) , 
\end{equation}
where $( k_\mu(z,s), k_6(z,s))$ is a local current which 
is gauge invariant under the $5+1$-dimensional gauge transformation.
Then we can see that $k_6(z,s)$ provides the desired 
local, gauge invariant field with the required properties.
In fact, from the $5+1$-dimensional gauge invariance and the fact that 
$q(z,s)$ is a linear functional of 
${\cal D}_s U_\mu \, {U_\mu}^{-1}$, we infer that 
$k_6(z,s)$ cannot depend on $A_6(z,s)$
and its $s$-dependence comes from that of the
link variable $U_\mu^s$. 
Then we can set
\begin{eqnarray}
\check Q_{\rm 5w} 
&\equiv &
\sum_z k_6(z,s=1)\\
&=& 
\sum_z \int_0^1 ds \, 
\left.\left\{ 
\eta_\mu^a(z,s) J_\mu^a(z) \vert_{U_\mu^s}
- \partial_\mu^\ast k_\mu(z,s)\right\vert_{A_6=0}
\right\} ,
\end{eqnarray}
assuming $\sum_z k_6(z,s=0)= 0$. 

From the $5+1$-dimensional gauge invariance of $k_\mu(z,s)$ 
and the fact that $q(z,s)$ is a linear functional of 
${\cal D}_s U_\mu \, {U_\mu}^{-1}$, 
we also infer that 
$k_\mu(z,s)$ must be a linear functional of 
${\cal D}_s U_\mu \, {U_\mu}^{-1}$ written as
\begin{equation}
k_\mu(z,s)=\sum_w j_{\mu\nu}^a(z,s;w)   
\left\{ {\cal D}_s U_\nu(w,s) \, U_\nu(w,s)^{-1} \right\}^a .
\end{equation}
This in turn implies that 
\begin{equation}
\label{eq:gauge-invariance-of-kmu}
 \delta \left\{ k_\mu(z,s)\vert_{U_\mu^s,A_6=0} \right\}
+ \sum_w 
\frac{\delta k_\mu(z,s)}{\delta A_6(w,s)} \delta A_6(w,s) 
 = 0,
\end{equation}
where 
$\delta U_\mu^s \left\{U_\mu^s\right\}^{-1}= -D_\mu \omega(z,s)$ 
and $\delta A_6=[ \omega(z,s), A_6 ] - \partial_s \omega(z,s)$.
On the other hand, by setting the link variables $s$-independent 
in Eq.~(\ref{eq:5+1-dim-topological-field-trivial}), we obtain
\begin{equation}
-\left\{D_\mu  A_6(z,s)\right\}^a 
\, J_\mu^a(z)\vert_{U_\mu}  
=
\partial_\mu^\ast \, 
\left\{ 
\sum_w 
\left. \frac{\delta k_\mu(z,s)}{\delta A_6(w,s)}  A_6(w,s) 
\right\vert_{U_\mu} 
\right\}.
\end{equation}
By setting $A_6(z,s)=-\partial_s \omega(z,s)$ in this equation 
and using Eq.~(\ref{eq:gauge-invariance-of-kmu}),
we obtain 
\begin{equation}
\label{eq:condition-for-gauge-invariance-true}
-\left\{D_\mu  \partial_s \omega(z,s)\right\}^a 
\, J_\mu^a(z)\vert_{U_\mu^s}  
=\partial_\mu^\ast \, \delta k_\mu(z)\vert_{U_\mu^s,A_6=0} .
\end{equation}
Thus Eq.~(\ref{eq:condition-for-gauge-invariance}) is indeed satisfied.

In this manner, we can reconstruct the Chern-Simons term 
with the required properties for all contractible loops,
through the cohomology problem with the 
topological field Eq.~(\ref{eq:5+1-dim-topological-field-1}).

\subsection{The ansatz for non-contractible loops}

When the topological field 
Eq.~(\ref{eq:5+1-dim-topological-field-1}) 
is shown to be cohomologically
trivial, the remaining issue is to show the condition 
Eq.~(\ref{eq:CS-local-field}) for all possible non-contractible loops.  
One may try $k_6(z,s)$ as the anzatz for the Chern-Simons term.
\begin{equation}
\check Q_{\rm 5w} = \sum_z k_6(z,s=1).
\end{equation}
The question is then to show 
\begin{equation}
\label{eq:integrability-condition-for-non-contractible-loops}
\exp(i Q_{\rm 5w}) = \exp(i \check Q_{\rm 5w})
\end{equation}
for all non-contractible loops. 

This problem requires first to figure out the topological structure 
of the space of the gauge fields in consideration,
which is constrained by the admissibility condition.
So far, the topological structure of the space of 
the admissible gauge field is known only for abelian 
gauge theories \cite{abelian-chiral-gauge-theory}.
In this case, 
it is indeed possible to show that the topological field 
Eq.~(\ref{eq:5+1-dim-topological-field-1}) is 
cohomologically trivial and 
to establish 
Eq.~(\ref{eq:integrability-condition-for-non-contractible-loops})
for all loops in the space of the gauge field,
as shown in the approach of the Ginsparg-Wilson relation 
\cite{abelian-chiral-gauge-theory}.

\subsection{Some results in the infinite four-dimensional volume}

In the infinite four-dimensional volume, the cohomology
problem in 5+1-dimensional space
defined in the previous subsections 
can be solved in certain cases. 
(cf. \cite{topology-on-the-lattice,
noncomutative-brs-cohomology,cohomology-non-abelian,perturbation-theory})
In this subsection, we describe how to construct 
the local counter terms  for the theories
in the infinite four-dimensional volume.

\subsubsection{Abelian chiral gauge theories}

In the abelian gauge theories, the lattice Chern-Simons current
is a gauge-invariant conserved current, which is a  local 
functional of the gauge fields:
\begin{equation}
\delta J_\mu(z) = 0, \qquad  \partial_\mu^\ast J_\mu(z) = 0.
\end{equation}
With these two conditions
we can directly apply the cohomological method 
using the 
Poincar\'e lemma on the lattice 
\cite{topology-on-the-lattice,noncomutative-brs-cohomology}
to $J_\mu(z)$, to obtain 
\begin{eqnarray}
\label{eq:CS-current-structure-abelian}
J_\mu(z) &=&
\alpha_\mu 
+
\beta_{\mu\nu\sigma} F_{\nu\sigma}(z-\hat\nu-\hat\sigma)
\nonumber\\
&&
\quad \ +\gamma \, \epsilon_{\mu\nu\sigma\rho\tau}
F_{\nu\sigma}(z-\hat\nu-\hat\sigma) 
F_{\rho\tau}(z-\hat\nu-\hat\sigma-\hat\rho-\hat\tau) \nonumber\\
&& \qquad \ + \partial_\nu^\ast \chi_{\nu\mu}(z)   ,
\end{eqnarray}
where $\chi_{\nu \mu}(z)$ is a gauge-invariant and local 
anti-symmetric tensor field and
\begin{equation}
  F_{\mu\nu}(z) 
= \frac{1}{i} \ln \left\{ U_\mu(z)U_\nu(z+\hat\mu)
U_\mu^{-1} (z+\hat\nu)U_\mu^{-1}(z) \right\}.
\end{equation}
From the lattice symmetries, we infer that $\alpha_\mu$ and
$\beta_{\mu\nu\rho}$ vanish identically.
Through explicit calculations in the weak coupling expansion,
we can also verify that $\alpha_\mu$ and $\beta_{\mu\nu\rho}$ 
vanish identically and $\gamma=-\frac{1}{32\pi^2}$ \cite{aoyama-kikukawa}.
From this result, we can reconstruct the lattice
Chern-Simons term as\footnote{Here 
$A_\mu(z)$ is the vector potential 
which represents the original admissible (five-dim.)
link variable $U_\mu(z)$ and the field strength as 
follows \cite{topology-on-the-lattice}:
\begin{equation}
  U_\mu(z) = e^{i A_\mu(z)}, \qquad
 \vert A_\mu(z) \vert \le \pi(1+10 \left\Arrowvert z \right\Arrowvert)
\end{equation}
\begin{equation}
F_{\mu\nu}(z) = \partial_\mu A_\nu(z) -   \partial_\nu A_\mu(z) .
\end{equation}
This vector potential itself is not a local functional 
of the original link variable $U_\mu(z)$, but
its local, gauge invariant functional becomes a local
functional of $U_\mu(z)$.
} 
\begin{eqnarray}
Q_{\rm 5w}
&=& \sum_z
\int_0^1 ds \, A_\mu(z) \left. J_\mu(z)\right\vert_{A\rightarrow sA}
\qquad (U_\mu(z;s)=e^{i s A_\mu(z)})
\nonumber\\
&=& 
\sum_z \frac{1}{3} \gamma \,
\epsilon_{\mu\nu\sigma\rho\tau} \, 
A_\mu(z)
F_{\nu\sigma}(z-\hat\nu-\hat\sigma) 
F_{\rho\tau}(z-\hat\nu-\hat\sigma-\hat\rho-\hat\tau) 
\nonumber\\
&& 
+ \sum_z \partial_\nu^\ast \hat \chi_{\nu\mu}(z)  A_\mu(z),
\end{eqnarray}
where 
\begin{equation}
\hat \chi_{\nu\mu}(z) = \int_0^1 ds 
\left. \chi_{\nu\mu}(z) \right\vert_{A\rightarrow s A}  .
\end{equation}

If we consider an anomaly free abelian chiral gauge 
theory, the charges of the Weyl fermions
should satisfy the condition 
\begin{equation}
  \sum_\alpha e_\alpha^3 = 0.
\end{equation}
Then, by rescaling the gauge field as $A_\mu \rightarrow e_\alpha
A_\mu$ in each Weyl fermion contributions, we can see that 
the first term of the lattice Chern-Simons term
vanishes identically. As to the second term, we may add
the following total divergence term without affecting 
the lattice Chern-Simons term: 
\begin{equation}
  - \sum_z \partial_\nu^\ast \left( \hat \chi_{\nu\mu}(z) A_\mu(z)\right).
\end{equation}
Then we obtain the local expression of the lattice Chern-Simons term
(the local counter term) which is manifestly gauge invariant:
\begin{equation}
  Q_{\rm 5w}= \sum_z \frac{1}{2} \hat \chi_{\nu\mu}(z-\hat \nu) 
F_{\nu\mu}(z-\hat \nu).
\end{equation}

\subsubsection{Non-abelian chiral gauge theories 
in lattice perturbation theory}

In the anomaly-free non-abelian chiral gauge theories, 
it is possible to show that 
the topological field Eq.~(\ref{eq:5+1-dim-topological-field-1})
is cohomologically trivial
to any orders of the lattice perturbation theory,
as shown in the approach based on the Ginsparg-Wilson relation 
\cite{cohomology-non-abelian,perturbation-theory}.
In particular, one can construct the local field
of the Chern-Simons term 
directly from the lattice Chern-Simons current, 
to any orders of the perturbation theory.

In the lattice perturbation theory, 
non-abelian lattice gauge fields 
are treated in the expansion of the gauge coupling constantant,
\begin{equation}
  U_\mu(z) = 1 + \sum_{l=1}^\infty \left( i g A_\mu(z) \right)^l.
\end{equation}
Accordingly, the local field of the lattice Chern-Simons term
may be assumed to have the expansion in the gauge coupling 
constant:
\begin{equation}
  Q_{\rm 5w}= \sum_z \sum_{l=1}^\infty q_{\rm 5w}^{(l)}(z)
\end{equation}
Let us assume that the local fields
$q_{\rm 5w}(z)$ are constructed 
to the order $l=n$ and consider how to 
construct the local field of the order $l=n+1$. 

Since the Chern-Simons term should produce the Chern-Simons current
under the local variation of the gauge field, 
\begin{equation}
  U_\mu(z) \longrightarrow U_\mu(z) + \eta_\mu(z) U_\mu(z), 
\end{equation}
we should have
\begin{equation}
\sum_{l=1}^\infty \delta_\eta  q_{\rm 5w}^{(l)}(z)
= \eta_\mu^a(z) J_\mu^a(z).
\end{equation}
Then we may expand the variation of the Chern-Simons current 
which is subtracted 
by the local fields up to the order $l=n$:
\begin{equation}
\eta_\mu^a(z) J_\mu^a(z) 
- \sum_{l=1}^n \delta_\eta q_{5w}^{(l)} (z)
= \eta_\mu^a(z) \check J_\mu^{a(n)}(z)  + {\cal O}(g^{n+2}).
\end{equation}
The leading term in this expansion,
$\check J_\mu^{a(n)}(z)$, is 
a local field of the vector potential of order $l=n$.
Since the l.h.s. is gauge invariant up to ${\cal O}(g^n)$, 
$\check J_\mu^{a(n)}(z)$ is invariant under the linearized 
gauge transformation,
\begin{equation}
  A_\mu^a(z) \rightarrow A_\mu^a(z) + \partial_\mu \omega^a(z)
\end{equation}
and the global gauge transformation. It also  
satisfies the conservation law,
\begin{equation}
 \partial_\mu^\ast \check J_\mu^{a (n)}(z) = 0.
\end{equation}
From these conditions, 
we can directly apply the cohomological method 
for the abelian gauge theories 
\cite{topology-on-the-lattice,noncomutative-brs-cohomology}
to $\check J^{a(n)}_\mu(z)$, to obtain 
\begin{equation}
\check J_\mu^{a (n)}(z)
= 
\partial_\nu^\ast \chi_{\nu \mu}^{a(n)}(z) , \qquad (n \not = 2).
\end{equation}
where $\chi_{\nu \mu}^{a(n)}(z)$ is a local field which is 
invariant under the linearized gauge transformation and 
the global gauge transformation. 
When and only when $n=2$, 
a cohomologically non-trivial term can appear as 
\begin{eqnarray}
\check J_\mu^{a (2)}(z)
&=&
d ^{abc}
\gamma \, \epsilon_{\mu \nu \rho \sigma \tau} 
  F_{\nu \rho}^b (z-\hat \nu -\hat \rho)   
F_{\sigma \tau}^c (z-\hat \nu -\hat \rho-\hat \sigma - \hat \tau),
\nonumber\\
&+& \partial_\nu^\ast \chi_{\nu \mu}^{a(2)}(z),
\end{eqnarray}
which vanishes identically by the anomaly-free condition 
\begin{equation}
\sum d^{abc}=\sum {\rm Tr}\left( T^a \left\{T^b, T^c\right\} \right) = 0.
\end{equation}

From this result, we can obtain the local field
of order $l=n+1$,
which reproduces $\check J_\mu^{a (n)}(z)$ under the local variation 
of the gauge field, as  
\begin{equation}
  \check q_{\rm 5w}^{(n+1)}(z) = 
\frac{1}{2} \check \chi_{\nu\mu}^{a(n)}(z-\hat\nu)
  \check F^a_{\nu\mu}(z-\hat\nu), 
\end{equation}
where
\begin{equation}
  \check F_{\nu\mu}^a(z)= \partial_\nu  A_\mu^a(z) 
                         -\partial_\mu  A_\nu^a(z) 
\end{equation}
and
\begin{equation}
   \check \chi_{\nu \mu}^{a(n)}(z)
 = \int_0^1 d s  \left. \chi_{\nu \mu}^{a(1)}(z) \right\vert_{A\rightarrow s A}
 = \frac{1}{n+1} \, \chi_{\nu \mu}^{a(n)}(z).
\end{equation}
Here we have used the fact that 
$\chi_{\nu \mu}^{a(n)}(z)$ has the following structure
in the vector potentials:
\begin{eqnarray}
\label{eq:chi-structure}
\chi_{\nu \mu}^{a(n)}(z)
&=&\sum
\chi_{\nu \mu}^{a (n);a_1 a_2\cdots a_{n}}
(z;z_1,z_2,\cdots,z_{n})_{\mu_1 \mu_2 \cdots \mu_{n}} \times
\nonumber\\
&& \qquad \qquad 
A_{\mu_1}^{a_1}(z_1) A_{\mu_2}^{a_2}(z_2) \cdots 
A_{\mu_{n}}^{a_{n}}(z_{n}).
\end{eqnarray}

The above field $\check q_{\rm 5w}^{(n+1)} (z)$
is invariant under the linearized gauge transformation and 
the global gauge transformation.
Next step is to construct $q_{\rm 5w}^{(n+1)} (z)$
which is gauge invariant 
under the full non-abelian gauge transformation, while its leading term 
in the expansion of the gauge coupling constant remains to coincide
with $\check q_{\rm 5w}^{(n+1)} (z)$.
For this purpose, we follow the method adopted in \cite{perturbation-theory}.
Namely, this step can be achieved by the replacement of the field 
strength as 
\begin{equation}
  \check F_{\nu\mu}^a(z) \longrightarrow
 \frac{2}{a} 
{\rm Tr} \left\{ 
T^a 
\left[ 1 - U_{\nu\mu}(z) \right]\right\},
\end{equation}
and by the replacement of the vector potentials
in Eq.~(\ref{eq:chi-structure}) as
\begin{equation}
A_\mu^a(z_k ) \longrightarrow 
\hat A_\mu^a(z,z_k) = 
\frac{2}{a} {\rm Tr} \left\{ 
T^a 
\left[ 1 - W(z,z_k) U_\mu(z_k) W(z,z_k+ \hat \mu)^{-1} \right]
\right\},
\end{equation}
where $W(z,z_k)$ is defined as the ordered product 
of the link variables from $z_k$ to $z$ along the shortest path
that first goes in direction $1$, then direction $2$, and so on. 
Since it coincides with the original vector potential 
up to the linearized gauge transformation 
in the expansion of the gauge coupling constant as
\begin{equation}
\hat A_\mu^a(z,z_k) = g \left\{ A_\mu(z_k) + \partial^{z_k}_\mu \omega(z,z_k) 
\right\} + {\cal O}(g^2) ,
\end{equation}
where $\omega(z,z_k)$ is the oriented line sum of the 
gauge potential from $z_k$ to $z$ along the same path as defined for
$W(z,z_k)$, the leading term of 
$q_{\rm 5w}^{(n+1)} (z)$ so defined actually
coincides with $\check q_{\rm 5w}^{(n+1)} (z)$.

\section{Connection to the lattice theory of Weyl fermion 
based on the Ginsparg-Wilson relation}
\label{sec:connection-dwf-and-GW}
\reseteqnum

In this section, we discuss the connection 
of the chiral domain wall fermion discussed so far 
to
the gauge-invariant construction of chiral gauge theories
based on the Ginsparg-Wilson 
relation \cite{abelian-chiral-gauge-theory,nonabelian-chiral-gauge-theory}.
We will establish an identity relating the 
partition function of the domain wall fermion 
(subtracted and with the local counter term)
and the partition function of the Weyl fermion defined with 
the overlap Dirac operator satisfying the Ginsparg-Wilson relation.

\subsection{Weyl fermion defined through
the overlap Dirac operator}

The lattice Dirac fermion theory defined with the Dirac operator
which satisfies the Ginsparg-Wilson relation
\begin{equation}
\gamma_5 D + D \hat \gamma_5 = 0, 
\qquad \hat \gamma_5 = \gamma_5(1-2 D) , %= - \frac{H}{\sqrt{H^2}}.
\end{equation}
possesses the exact symmetry under the chiral transformation
\begin{equation}
  \delta \psi(x) = \hat \gamma_5 \psi(x), \qquad
  \delta \bar \psi(x) = \bar \psi(x) \gamma_5 .
\end{equation}
Based on this exact chiral symmetry, 
the lattice Weyl fermion can be defined
by imposing the constraint with the 
chiral operators $\hat \gamma_5$ and $\gamma_5$,
\begin{equation}
  \hat \gamma_5 \psi_R(x) = + \psi_R(x) , \quad
  \bar \psi_R(x) \gamma_5 = - \bar \psi_R(x).
\end{equation}
By introducing the orthnormal basis for the Weyl fermion, 
$\left\{ v_i(x) \vert i=1,2,\cdots \right\}$ 
and
$\left\{ \bar v_k(x) \vert k=1,2,\cdots \right\}$\footnote{
 $(v_i,v_j)$ is the inner product of the spinor field defined by
\begin{equation}
(v_i,v_j) = \sum_x v_i(x)^\dagger v_j(x) ,
\end{equation}
} 
as 
\begin{equation}
  \hat \gamma_5 v_i(x) = + v_i(x), \qquad (v_i,v_j)=\delta_{ij},
\end{equation}
\begin{equation}
  \bar v_k(x) \gamma_5  = - \bar v_k(x), \qquad (\bar v_k,\bar v_l)
=\delta_{kl},
\end{equation}
the functional measure of the Weyl fermion can be set up and 
the path-integral formula of the partition function can be defined.
\begin{equation}
\label{eq:path-integral-partition-function-Weyl-fermion}
Z_w = \int {\cal D}[\psi_R]{\cal D}[\bar \psi_R]
\, e^{-\sum_x \bar \psi_R(x) D \psi_R(x) }
= \det (\bar v_k, D v_j).
\end{equation}

The choice of the basis 
$\left\{ v_i(x) \vert i=1,2,\cdots \right\}$ 
may be different by a unitary transformation, 
\begin{equation}
v_i(x)  \longrightarrow \tilde v_i(x)=\sum_j v_j(x) Q_{ji} ,
\end{equation}
which dependents on the gauge field in general 
because $\hat \gamma_5$ does so.
Then the measure 
is changed by the phase factor 
\begin{equation}
{\cal D}[\psi_R]{\cal D}[\bar \psi_R]
\longrightarrow 
{\cal D}[\psi_R]{\cal D}[\bar \psi_R]\,
\det \left\{ Q_{ji} \right\} .
\end{equation}
Accordingly, the partition function is changed by
\begin{equation}
\det (\bar v_k, D v_j) 
\longrightarrow 
\det (\bar v_k, D \tilde v_j)  
= \det (\bar v_k, D v_j) \det \left\{ Q_{ji} \right\}.
\end{equation}
In the gauge-invariant construction of the chiral gauge 
theories based on the Ginsparg-Wilson relation 
\cite{abelian-chiral-gauge-theory,nonabelian-chiral-gauge-theory},
a general method to fix the phase of the functional measure
(and the partition function) has been described so that it satisfies 
the requirements of the smoothness, the locality and the gauge invariance.

Our target in this paper is the Weyl fermion theory which is 
defined with the overlap Dirac operator 
satisfying the Ginsparg-Wilson 
relation \cite{overlap-D,overlap-D-GW-relation}.
The overlap Dirac operator, $D$, is given by the 
explicit formula \cite{overlap-D}
\begin{equation}
D = \frac{1}{2}\left( 1 + \gamma_5 \frac{H}{\sqrt{H}} \right).
\end{equation}
Here $H$ is chosen as the hermitian operator 
obtained through the transfer matrix of the 
five-dimensional Wilson-Dirac fermion.\footnote{
Bori\c ci has pointed out that 
the transfer matrix 
can be expressed by a 
simpler four-dimensional hermitian matrix as
\[ T=\frac{1+{\cal H}}{1-{\cal H}} , \qquad
{\cal H} = \gamma_5 (D_{\rm w}-m_0)
\frac{1}{1+a_5(D_{\rm w}-m_0)},\]
which leads to the same spectrum asymmetry operator,
${\cal H}/\sqrt{{\cal H}^2}=H/\sqrt{H^2}$ \cite{borici}.
}
\begin{equation}
H = - \ln T .
\end{equation}
In this case $\hat \gamma_5$ is given by 
the spectrum asymmetry of $H$,
\begin{equation}
\hat \gamma_5 = \gamma_5(1-2 D) = - \frac{H}{\sqrt{H^2}},
\end{equation}
and the chiral basis, 
$\left\{ v_i(x) \vert i=1,2,\cdots \right\}$, 
can be chosen as the eigenvectors of $H$ belonging to the negative 
eigenvalues (up to the global phase choice).
The partition function resulted from the path-integral formula
Eq.~(\ref{eq:path-integral-partition-function-Weyl-fermion}),
reproduces the overlap formula of the chiral determinant \cite{overlap}:
\begin{equation}
\label{eq:partition-function-overlap}
\det (\bar v_k, D v_j)  = \det (\bar v_k,v_j) .
\end{equation}
Here the phase of the chiral basis can be chosen following the 
gauge-invariant method of 
\cite{abelian-chiral-gauge-theory,nonabelian-chiral-gauge-theory}
in anomaly-free chiral gauge theories.

\subsection{Gauge-invariant partition function of Weyl fermions 
from the domain wall fermion }

We now argue the connection of 
the (subtracted) partition function
of the domain wall fermion with the local counter term,
\begin{equation}
\label{eq:dwf-partition-function-subtracted-CS}
\lim_{N\rightarrow \infty}
\frac{ \left. \det \left(D_{\rm 5w}-m_0\right) \right\vert_{\rm Dir.}^{c_1} }
{ \left\vert \left. \det \left(D_{\rm 5w}-m_0\right) 
\right\vert_{\rm AP} ^{c_1+(-c_1)}
  \right\vert^{\frac{1}{2}} }
\, e^{-iC_{\rm 5w}(c_1)} 
\end{equation}
to the partition function of the Weyl fermion given by
Eq.~(\ref{eq:partition-function-overlap}).\footnote{
It may be worth while to recall the situation in
the vector-like case.
In the vector-like theories \cite{boundary-fermion}, 
%\footnote{The author refer the reader
%to \cite{vranas-lat00,creutz-review} 
%for recent reviews on the application of
%domain wall fermion.} % to  vector-like theories.}
the connection between the domain wall fermion and 
the overlap Dirac operator 
%satisfying the Ginsparg-Wilson relation  
is directly seen in the fact that  the determinant of
domain wall fermion (subject to Dirichlet boundary condition
in the fifth dimension) is
factorized into  four-dimensional part for the low-lying massless
mode  and five-dimensional part for the remaining massive 
modes \cite{truncated-overlap,vranas-pauli-villars}:
\begin{equation}
\label{eq:domain-wall-fermion-vs-D}
 \lim_{N \rightarrow \infty} 
\frac{
\left. \det \left( D_{\rm 5w} - m_0 \right)\right\vert_{\rm Dir.}
}
{
\left. \det \left( D_{\rm 5w} - m_0 \right)\right\vert_{\rm AP} 
} 
=  \det D .
\end{equation}
In the l.h.s., $N$ is the size of the fifth dimension
and 
the contribution of the massive modes is factorized in
the determinant of the five-dimensional Wilson fermion
subject to the anti-periodic boundary condition in the 
fifth dimension \cite{vranas-pauli-villars}.
}

\subsubsection{Partition functions of the 5-dim. Wilson fermions}

The partition functions of 
the five-dimensional Wilson fermions with 
the Dirichlet boundary condition and 
with the anti-periodic boundary condition 
can be expressed explicitly in terms of the 
transfer matrix, 
as shown in the appendix~\ref{app:determinant-of-DWF}.
The results are given as follows:
\begin{eqnarray}
\label{eq:partition-function-chiral-DWF}
&& \left.  
\det \left(D_{\rm 5w}-m_0 \right) \right\vert_{\rm Dir.}^{c_1}
\nonumber\\
&&\quad = 
\det 
\left( 
{\scriptstyle
P_R + P_L \, 
          T_1^{(N-\Delta)} 
          \biggl\{ U_5^{-1} \prod_{c_1}%_{t=-\Delta+1}^{\Delta} 
                   T_t U_{5,t-1}^{-1} \biggr\} \, \, 
          T_0^{N-\Delta}
}
\right)  \nonumber\\
&& \qquad 
\times 
\det\left(
{\scriptstyle 
P_R + P_L \prod_{c_1} U_{5,t} 
}
\right)
\, \prod_{t=-N+1}^{N} N_t ,
\\
&&\nonumber\\
\label{eq:partition-function-5DWF-AP}
&& \left.  
\det \left(D_{\rm 5w}-m_0 \right) \right\vert_{\rm AP}^{c_1+(-c_0)}
\nonumber\\
&&\quad = 
\det 
\left( 
{\scriptstyle
1 + \, 
          T_0^{N-\Delta}
          \biggl\{ U_5^{-1} \prod_{-c_0}%_{t=-\Delta+1}^{\Delta} 
                   T_t U_{5,t-1}^{-1} \biggr\} \, \, 
          T_1^{2(N-\Delta)} 
          \biggl\{ U_5^{-1} \prod_{c_1}%_{t=-\Delta+1}^{\Delta} 
                   T_t U_{5,t-1}^{-1} \biggr\} \, \, 
          T_0^{N-\Delta}
}
\right)  \nonumber\\
&& \qquad 
\times 
\det\left(
{\scriptstyle 
P_R + P_L \prod_{c_1+(-c_0)} U_{5,t}
}
\right)
\, \prod_{t=-N+1}^{3N} N_t ,
\end{eqnarray}
where $T_t$ is the transfer matrix 
%of the five-dimensional Wilson-Dirac fermion
with $U_k(x,t)$, which explicit form is given 
in appendix~\ref{app:transfer-matrix},
and $N_t = \det\left(P_L + P_R B_t \right)$.
The subscript $0$ and $1$ denote
the quantities with $U_k^0(x)$ and $U_k^1(x)$, respectively.
$U_5(x,t)$ appears in between the product of the transfer matrices
so that the five-dimensional gauge covariance is maintained. Note 
that for abelian gauge group, the extra phase factor, which consists of
the product of $U_5(x,t)^{-1}$, appears.

In these formula, the dominant term 
in $T_0^{(N-\Delta)}$ in the limit $N \rightarrow \infty$
is estimated as follows:
\begin{eqnarray}
\label{eq:estimation-T-to-infinity}
T_0^{(N-\Delta)} &=& P_0 T_0^{(N-\Delta)}
                   +(1-P_0)T_0^{(N-\Delta)}
\nonumber\\
&=& 
 \sum_i v_i^0 \otimes {v_i^0}^\dagger e^{(N-\Delta) | \lambda_i ^0| }
+ {\cal O}( e^{-(N-\Delta) \lambda_+^0 } ) ,
\end{eqnarray}
where the projection operator $P_0$ is introduced by
\begin{equation}
  P_0 = \frac{1}{2}\left(1-\frac{H_0}{\sqrt{H^2_0}} \right), \quad
 H_0 = - \ln T_0, 
\end{equation}
and $\{v_i^0(x)\}$ are chosen as
eigenfunctions of $H_0=-\ln T_0$ belonging to
the negative eigenvalues $\lambda_i^0$, while
$\lambda_+^0$ is the smallest positive eigenvalue of $H_0$.
Similar estimation holds for $T_1^{(N-\Delta)}$.
With these results, we can infer that
in the limit $N \rightarrow \infty$
\begin{eqnarray}
\label{eq:partition-function-dwf-limit}
&&
\lim_{N \rightarrow \infty}
\frac{\left.  \det \left(D_{\rm 5w}-m_0 \right) 
\right\vert_{\rm Dir.}^{c_1}}
    { 
      \biggl\{ \prod_{t=-N+1}^N N_t \biggr\}
       \, N_1^{N-\Delta} 
       \, N_0^{N-\Delta} 
}
\nonumber\\
&&=
\det 
\left( P_R + P_L \, 
       P_1 %  T_1^{T-\Delta} 
          \biggl\{ 
U_{5,\Delta}^{-1}  \prod_{c_1} T_t U_{5,t-1}^{-1} 
\biggr\}
\, \, 
       P_0 % T_0^{T-\Delta}
\right) \nonumber\\
&& \qquad \qquad \qquad \qquad
\times
\det\left(P_R + P_L \prod_{c_1}U_{5,t} \right) ,
\\
&& \nonumber\\
\label{eq:partition-function-5d-Wilson-ap-limit}
&&
\lim_{N \rightarrow \infty}
\frac{\left.  \det \left(D_{\rm 5w}-m_0 \right) 
\right\vert_{\rm AP}^{c_1+(-c_0)}}
    { 
      \biggl\{ \prod_{t=-N+1}^{3N} N_t \biggr\}
       \, N_1^{2(N-\Delta)} 
       \, N_0^{2(N-\Delta)} 
}
\nonumber\\
&&=
\det 
\left( 1-P_0 + P_0 \, 
       %  T_1^{T-\Delta} 
          \biggl\{ U_5^{-1} \prod_{-c_0}%_{t=-\Delta+1}^{\Delta} 
                   T_t U_{5,t-1}^{-1} \biggr\} \, \, 
P_1 
          \biggl\{ U_5^{-1} \prod_{c_1}%_{t=-\Delta+1}^{\Delta} 
                   T_t U_{5,t-1}^{-1} \biggr\} \, \, 
P_0
\right) 
\nonumber\\
&& \qquad \qquad \qquad \qquad 
\times
\det\left(P_R + P_L \prod_{c_1+(-c_0)}U_{5,t} \right) ,
\end{eqnarray}
where
$N_1=\det\left(1-P_1 + P_1 T_1\right)$ and
$N_0=\det\left(1-P_0 + P_0 T_0\right)$.
(See appendix~\ref{app:partition-function-limit} 
for the derivation of these results. )

From Eqs.~(\ref{eq:partition-function-dwf-limit}) and 
(\ref{eq:partition-function-5d-Wilson-ap-limit}), it follows 
immediately that 
\begin{eqnarray}
\label{eq:partition-function-dwf-subtracted-CS-limit}
&&\lim_{N\rightarrow \infty}
\frac{ \left. \det \left(D_{\rm 5w}-m_0\right) \right\vert_{\rm Dir.}^{c_1} }
{ \left\vert \left. \det \left(D_{\rm 5w}-m_0\right) 
\right\vert_{\rm AP} ^{c_1+(-c_1)}
  \right\vert^{\frac{1}{2}} }
\, e^{-iC_{\rm 5w}(c_1)} 
\nonumber\\
&& =
\frac{\scriptstyle
\det 
\left( P_R + P_L \, 
       P_1 %  T_1^{T-\Delta} 
          \biggl\{ 
U_{5,\Delta}^{-1} \prod_{c_1} T_t U_{5,t-1}^{-1}
\biggr\}
\, \, 
       P_0 % T_0^{T-\Delta}
\right) \times \det\left(P_R + P_L \prod_{c_1}U_{5,t} \right)}
{\scriptstyle
\left\vert \det 
\left( 1-P_0 + P_0 \, 
       %  T_1^{T-\Delta} 
          \biggl\{ U_5^{-1} \prod_{-c_1}%_{t=-\Delta+1}^{\Delta} 
                   T_t U_{5,t-1}^{-1} \biggr\} \, \, 
P_1 
          \biggl\{ U_5^{-1} \prod_{c_1}%_{t=-\Delta+1}^{\Delta} 
                   T_t U_{5,t-1}^{-1} \biggr\} \, \, 
P_0
\right) \right\vert^{\frac{1}{2}}
}
\, e^{-iC_{\rm 5w}(c_1)} .
\nonumber\\
\end{eqnarray}

\subsubsection{Factorization of the partition function of 
the domain wall fermion}

Now we introduce
the chiral basis $\{v_i^0(x)\}$
associated with the gauge fields $U_k^0(x)$ as
\begin{equation}
  P_0 v_i^0(x) = v_i^0(x),  \quad
(v_i^0,v_j^0) = \delta_{ij}
\qquad (i,j= 1, 2, \cdots ).
\end{equation}
$\{v_i^0(x)\}$ may be chosen as 
the eigenfunctions of $H_0=-\ln T_0$ belonging to
the negative eigenvalues.
Similarly, we introduce the chiral basis $\{v_i^1\}$  
associated with the gauge fields $U_k^1(x)$.
We also introduce the chiral basis for the anti-field $\{\bar v_k\}$ as
\begin{equation}
  \bar v_k(x) P_L = \bar v_k(x) \qquad (k = 1, 2,\cdots ).
\end{equation}

In terms of these chiral bases, the formula 
Eq.~(\ref{eq:partition-function-dwf-subtracted-CS-limit}) 
of the (subtracted) partition function of the domain wall fermion
with the local counter term can be rewritten further as 
\begin{eqnarray}
\label{eq:dwf-partition-function-subtracted-basis}
&& \lim_{N\rightarrow \infty}
\frac{ \left. \det \left(D_{\rm 5w}-m_0\right) \right\vert_{\rm
    Dir.}^{c_1}}
{ \left\vert \left. \det \left(D_{\rm 5w}-m_0\right) 
\right\vert_{\rm AP} ^{c_1+(-c_1)}
  \right\vert^{\frac{1}{2}} }
\, e^{-iC_{\rm 5w}(c_1)}  
\nonumber\\
&& =
\det \left(\bar v_k, v_i^1\right) \, e^{i\phi(c_1)} \,  
\det \left(\bar v_k, v_j^0\right)^\ast \,
e^{-iC_{\rm 5w}(c_1)}, 
\end{eqnarray}
where
\begin{equation}
\label{eq:phase-due-to-interpolation}
e^{i \phi(c_1) }= 
\frac{ \scriptstyle
 \det \left( v_i^1, \biggl\{ 
U_{5,\Delta}^{-1} \prod_{t=-\Delta+1}^{\Delta} T_t U_{5,t-1}^{-1} 
    \biggr\} \,  v_j ^0\right)}
{\scriptstyle
\left\vert  \det \left( v_i^1, \biggl\{ 
U_{5,\Delta}^{-1} \prod_{t=-\Delta+1}^{\Delta} T_t U_{5,t-1}^{-1} 
    \biggr\} \,  v_j^0 \right)\right\vert}
\,
\times
{\scriptstyle
\det\left(P_R + P_L \prod_{t=-\Delta+1}^{\Delta}U_{5,t} \right)}
.
\end{equation}
The first factor in the r.h.s. of 
Eq.~(\ref{eq:dwf-partition-function-subtracted-basis})
is nothing but the overlap formula, which gives   
the partition function of the right-handed Weyl fermion 
at $t=N$ coupled to the gauge field $U_k^1(x)$,
\begin{equation}
\det \left(\bar v_k, v_i^1\right)
=\det \left(\bar v_k, D \, v_i^1\right) .
\end{equation}
Similarly, the third factor 
in the r.h.s. of Eq.~(\ref{eq:dwf-partition-function-subtracted-basis})
reproduces the partition function of
the left-handed Weyl fermion at $t=-N+1$, which couples to 
the gauge field $U_k^0(x)$. 
On the other hand, the second factor in the r.h.s. of 
Eq.~(\ref{eq:dwf-partition-function-subtracted-basis}), 
which is the phase factor defined 
by Eq.~(\ref{eq:phase-due-to-interpolation}), 
\begin{equation}
e^{i\phi(c_1)}
\end{equation}
comes from the interpolation between the gauge 
fields $U_k^0(x)$ and $U_k^1(x)$.\footnote{In the original derivation 
of the overlap formula
in \cite{overlap},  this term was not considered because the gauge field
was assumed to be four-dimensional. It was also true in
the wave-guide model \cite{waveguide-kaplan,waveguide-jansen-etal}.} 
It depends on the choices of the chiral bases 
$\{v_i^0\}$, $\{v_i^1\}$ and the path
$c_1$, but  the path-dependence is to be compensated by the local 
counter term,
\begin{equation}
e^{-iC_{\rm 5w}(c_1)}.
\end{equation}

From these results, it is quite natural to choose 
the chiral basis of the Weyl fermion coupled to the gauge field
$U_k^1$  as follows:
\begin{equation}
\label{eq:choice-of-chiral-basis-from-dwf}
  v_i(x) = \left\{
       \begin{array}{ll}
          v_i^1(x) \, \,
e^{i \phi(c_1)} 
\, e^{-iC_{\rm 5w}(c_1)} %W(c_1)^{-1} 
& (i=1) \\
          v_i^1(x)            & (i \not = 1) 
       \end{array}
           \right.  .
\end{equation}
With this choice, 
the (subtracted) partition function of the domain wall 
fermion with the local counter term is factorized into
two chiral determinants:
\begin{equation}
\label{eq:partition-function-dwf-limit-factorized}
\lim_{N\rightarrow \infty}
\frac{ \left. \det \left(D_{\rm 5w}-m_0\right) \right\vert_{\rm Dir.}^{c_1} }
{ \left\vert \left. \det \left(D_{\rm 5w}-m_0\right) 
\right\vert_{\rm AP} ^{c_1+(-c_1)}
  \right\vert^{\frac{1}{2}} }
\, e^{-iC_{\rm 5w}(c_1)}
=\det\left(\bar v_k,D v_j \right)
  \det\left(\bar v_k,D v_j^0 \right)^\ast .
\end{equation}
The path-independence and the gauge invariance of 
$\det\left(\bar v_k,D v_j \right)$
are obvious from this identity.\footnote{
The complex phase part
of this identity can be regarded as the lattice counter part
of the relation between the $\eta$-invariant
and the effective action for the chiral 
fermions \cite{eta-invariant,ball-osbor,
atiyah-padoti-singer,domain-wall-ferimon-and-eta-invariant,
aoyama-kikukawa}.
Note also that this result 
is a generalization of Eq.~(\ref{eq:domain-wall-fermion-vs-D}) to 
the case of chiral gauge theories.  }

\subsection{The functional measure of the Weyl fermion
from the domain wall fermion}

The choice of the chiral basis 
in Eq.~(\ref{eq:choice-of-chiral-basis-from-dwf})
indeed leads to the functional measure of the Weyl fermion which is 
independent of the path of the interpolation.
To see this, we choose another path, say $c_2$, then we get
another basis,
\begin{equation}
\tilde  v_i(x) = \left\{
       \begin{array}{ll}
          v_i^1(x) \, \,
e^{i \phi(c_2)} 
\, e^{-iC_{\rm 5w}(c_2)} 
& (i=1) \\
          v_i^1(x)            & (i \not = 1) 
       \end{array}
           \right.  .
\end{equation}
These two bases are related by the unitary transformation $Q_{ij}$
\begin{equation}
\tilde v_i(x) = Q^{-1}_{ij} v_j(x), 
\end{equation}
which determinant turns out to be
\begin{equation}
\det Q =
e^{- i \phi(c_2)} 
\, e^{iC_{\rm 5w}(c_2)} 
\cdot
e^{i \phi(c_1)} 
\, e^{-iC_{\rm 5w}(c_1)}.
\end{equation}

But it is not difficult to see that 
the phase factors along the paths $c_1$ and $c_2$ multiply to give 
\begin{eqnarray}
\label{eq:integrability-condition}
e^{i \phi(-c_2)}  \cdot e^{i \phi(c_1)} 
&=& 
\frac{\scriptstyle 
\det\left(1-P_0 + P_0 \biggl\{ 
U_5^{-1} \prod_t^{-c_2} T_t U_{5,t-1}^{-1} 
    \biggr\} P_1 \biggl\{ 
U_5^{-1} \prod_t^{c_1} T_t U_{5,t-1}^{-1} 
    \biggr\} \right)}
{\scriptstyle
\left\vert \det\left(1-P_0 + P_0 \biggl\{ 
U_5^{-1} \prod_t^{-c_2} T_t U_{5,t-1}^{-1} 
    \biggr\} P_1 \biggl\{ U_5^{-1} \prod_t^{c_1} T_t U_{5,t-1}^{-1} 
    \biggr\} \right) \right\vert}
\nonumber\\
&& \qquad\qquad\qquad
\times
\frac{\scriptstyle
\det\left(P_R + P_L \prod_t^{c_1+(-c_2)} U_{5,t} \right)}
{\scriptstyle
\left\vert
\det\left(P_R + P_L \prod_t^{c_1+(-c_2)} U_{5,t} \right)
\right\vert}  \nonumber\\
&=&
\lim_{N\rightarrow \infty}
\frac{ \left. \det \left(D_{\rm 5w}-m_0\right) \right\vert_{AP}^{c_1+(-c_2)}}
{ \left\vert \left. \det \left(D_{\rm 5w}-m_0\right) 
\right\vert_{AP}^{c_1+(-c_2)}
  \right\vert } \nonumber\\
&=&e^{i Q_{\rm 5w}^{c_1+(-c_2)}} ,
\end{eqnarray}
and this identity can be regarded as the integrability condition
for the phase factor $\exp(i\phi(c_i))$.

Since we are assuming that the lattice Chern-Simons term can be 
decomposed into the two parts $C_{\rm 5w}^{c_1}$ and
$C_{\rm 5w}^{-c_2}$ locally and gauge-invariantly,
\begin{equation}
e^{i Q_{\rm 5w}^{c_1+(-c_2)}} = e^{i C_{\rm 5w}^{c_1}}\,
\cdot e^{-i C_{\rm 5w}^{c_2}},
\end{equation}
the determinant of the unitary transformation, $\det Q$, turns
out to be unity.
This holds for any two paths. Therefore the measure does not
depend on the path of the interpolation and is determined 
uniquely. 

We should note that 
from the integrability condition Eq.~(\ref{eq:integrability-condition})
and the expression for the (subtracted) partition function
Eq.~(\ref{eq:dwf-partition-function-subtracted-basis}), 
we can directly infer that
Eq.~(\ref{eq:dwf-partition-function-path-independence})
holds and the (subtracted) partition function with 
the local counter term is independent on the path of 
the interpolation for all topological sectors.

\subsection{The connection to the gauge-invariant 
construction based on the Ginsparg-Wilson relation}

Let us now look closely at the 
connection to the gauge-invariant construction 
by L\"uscher
\cite{abelian-chiral-gauge-theory,nonabelian-chiral-gauge-theory}.  
We note
first that our result 
Eq.~(\ref{eq:partition-function-dwf-limit-factorized}) should be compared
with the equation Eq.~(7.1) in \cite{nonabelian-chiral-gauge-theory}
\footnote{ Since we are considering the right-handed Weyl fermions,
instead of the left-handed Weyl fermions,
$P_R$ in the original equation Eq.~(7.1) of
\cite{nonabelian-chiral-gauge-theory}
is replaced by $P_L$.}:
\begin{equation}
\label{eq:partition-function-dwf-like}
\det \left(1-P_L + P_L D Q_1 D_0^\dagger
\right) W^{-1}=   \det (\bar v_k, D v_j)
                  \det (\bar v_k, D v^0_j)^\dagger ,
\end{equation}
where $Q_1$ is the evolution operator of the chiral  
projector
\begin{equation}
P_1 = Q_1 \, P_0 \,   {Q_1}^{-1},
\end{equation}
while $W$ is the Wilson line  defined by the line integral of 
the measure term current,
\begin{equation}
W=
e^{i \int_0^1 dt \, 
%\eta_k^a(x,t) 
\left\{\partial_t U_k(x,t) U_k(x,t)^{-1}\right\}^a
j_k^a(x,t) } ,
\end{equation}
where $t$ is the continuous parameter of the interpolation.

The determinant in the l.h.s. of Eq.~(\ref{eq:partition-function-dwf-like})
may be expressed in terms of the chiral bases $\{v^1_j\}$,
$\{v^0_j\}$ and $\{\bar v_k\}$ as 
\begin{equation}
\det \left(1-P_L + P_L D Q_1 D_0^\dagger \right) 
=\det(\bar v_k, D v_i^1) 
 \det(v_i, Q_1 v_j^0)
\det(\bar v_k, D v_j^0)^\ast .  
\end{equation}
Then we note the correspondence as 
\begin{equation}
  \det \left( v_i^1, Q_1 \,  v_j^0 \right) 
\quad  \Longleftrightarrow \quad
e^{i \phi(c_1)} =
\frac{
  \det \left( v_i^1, \biggl\{ 
\prod_{t=-\Delta+1}^{\Delta (c_1)} T_t 
    \biggr\} \,  v_j^0 \right) 
}{
\left\vert
  \det \left( v_i^1, \biggl\{ 
\prod_{t=-\Delta+1}^{\Delta (c_1)} T_t 
    \biggr\} \,  v_j^0 \right) 
\right\vert}  ,
\end{equation}
(in the gauge $U_5(z)=1$), which implies that in 
the domain wall fermion
the evolution of the basis of the Weyl fermion is 
realized by the successive multiplications of the 
transfer matrices, 
like time-development along the fifth 
dimension.\footnote{
We note that the use of the transfer matrix here
is in the same spirit as the use of the 
Hamiltonian for the evolution of the (second quantized) 
vacuum states of the overlap formalism, adopted in 
the adiabatic phase choice \cite{adiabatic-phase-chioce}.
Our result then provides a discretization method of the 
continuous evolution.}
For the loop in the space of the gauge fields,
this leads to the correspondence of the 
integrability conditions:
\begin{equation}
e^{-\int_0^1 dt \, 
\sum_i \left( v_i, \partial_t v_i \right) } 
= \det (1-P_0 + P_0 Q_1) \, \Longleftrightarrow \, 
  e^{i\phi(c_1)} \cdot e^{-i\phi(c_2)} = e^{i Q_{\rm 5w}^{c_1+(-c_2)}} .
\end{equation}

The above correspondences become exact in 
the continuum limit of the fifth dimension,
which should be taken as $a_5 \rightarrow 0 $ with $\Delta a_5$ fixed.
In taking the continuum limit,
one may smooth the interpolation further by 
replacing  $T_t$ with $T_t^{N_t}$ and take 
the limit $N_t \rightarrow \infty$ first as an intermediate step,
to get
\begin{equation}
\label{eq:continuum-limit-intermediate-step}
e^{i \phi(c_1)} =
\frac{
  \det \left( v_i^1, \biggl\{ 
%U_{5,\Delta} 
\prod_{t=-\Delta+1}^{\Delta (c_1)} P_t 
%U_{5,t-1}
    \biggr\} \,  v_j^0 \right) 
}{
\left\vert
  \det \left( v_i^1, \biggl\{ 
%U_{5,\Delta} 
\prod_{t=-\Delta+1}^{\Delta (c_1)} P_t 
%U_{5,t-1}
    \biggr\} \,  v_j^0 \right) 
\right\vert} .
\end{equation}
Then, using $P_t = Q_t P_0 {Q_t}^{-1} $, we can see 
that this becomes identical to 
$\det \left( v_i^1, Q_1 \,  v_j^0 \right)$
in the limit $a_5 \rightarrow 0 $ with $\Delta a_5$ fixed.

On the other hand, 
the dependence of the determinant in the l.h.s. 
of Eq.~(\ref{eq:partition-function-dwf-like}) 
on the path of the interpolation is compensated by 
the Wilson line $W$. 
Here the measure term 
$\sum_x \left\{\partial_t U_k(x,t) U_k(x,t)^{-1}\right\}^a j_k^a(x,t)$ 
is obtained from the 
topological field in $4+2$-dimensional space.
$W$ corresponds to 
the local counter term $\exp(i C_{\rm 5w}(c_1))$
in the domain wall fermion,
where the local field $k_6(x,t)$ is obtained from the 
topological field in $5+1$-dimensional space:
\begin{equation}
W=
e^{i \int_0^1 dt \, 
%\eta_k^a(x,t) 
\left\{\partial_t U_k(x,t) U_k(x,t)^{-1}\right\}^a
j_k^a(x,t) } 
\, \Longleftrightarrow \,
e^{iC_{\rm 5w}(c_1)}=
e^{i \sum_{t=-\infty}^{\infty}\sum_x k_6(x,t) } .
\end{equation}

In this respect, it is possible to see that
the topological field in $5+1$-dimensional space
reduces to 
the topological field in $4+2$-dimensional space introduced by
L\"uscher \cite{nonabelian-chiral-gauge-theory},
in the continuum limit.
For this, we recall that the $5+1$-dimensional topological 
field is 
defined from the local variation of the Chern-Simons term.
From Eq.~(\ref{eq:continuum-limit-intermediate-step})
and Eq.~(\ref{eq:integrability-condition}) 
the Chern-Simons term can be expressed as follows, 
in the same intermediate step as above:
\begin{eqnarray}
Q_{\rm 5w}^{c_1+(-c_2)}
&=& 
%\frac{
{\rm Im} \ln  \det \left( 1-P_0 + P_0 
\biggl\{ 
%U_{5,\Delta} 
\prod_{t=-\Delta+1}^{\Delta (-c_2)} P_t 
%U_{5,t-1}
\biggr\} \,  
P_1
\biggl\{ 
%U_{5,\Delta} 
\prod_{t=-\Delta+1}^{\Delta (c_1)} P_t 
%U_{5,t-1}
\biggr\} \,  
P_0 \right) . \nonumber\\
\end{eqnarray}
In this equation, one may express each $P_t$ 
using the chiral basis as 
$P_t = \sum_i v_i^t \otimes {v_i^t}^\dagger$ and
factorize the determinant. 
Then we consider the minimal deformation of the loop $c_1+(-c_2)$
at a certain point $t_0+1$ in the path $c_1$ as
\begin{equation}
  U_k(x,t_0+1) \equiv U_k^{s=0}(x,t_0+1)
\Longrightarrow 
  U_k(x,t_0+1)^\prime \equiv U_k^{s=\Delta s}(x,t_0+1)  ,
\end{equation}
and evaluate the variation of the Chern-Simons term, to obtain
\begin{eqnarray}
  \Delta Q_{\rm 5w}^{c_1+(-c_2)}
&=&
 {\rm Im} \ln \det \left( v_i^{t_0}, \{v_j^{t_0+1}\}^\prime \right) 
+{\rm Im} \ln \det \left( \{v_i^{t_0+1}\}^\prime, v_j^{t_0} \right) 
\nonumber\\
&-& 
{\rm Im} \ln \det \left( v_i^{t_0}, v_j^{t_0+1} \right) 
\ \quad -{\rm Im} \ln \det \left( v_i^{t_0+1}, v_j^{t_0} \right) 
\nonumber\\
&=& 
{\rm Im} \ln \det 
\left( 1-P^{t_0} 
     + P^{t_0} P^{t_0+1} P^{t_0+2} \{ P^{t_0+1} \}^\prime P^{t_0}
\right).
\nonumber\\
\end{eqnarray}
In the limit $\Delta s \rightarrow 0$ and $a_5 \rightarrow 0$, 
this variation reduces to 
\begin{equation}
i {\rm Tr} P \left[\partial_s P, \partial_5 P\right]  \, a_5 \, \Delta s .
\end{equation}
In the generic gauge $U_5(z)\not = 1$, this result reads
\begin{equation}
i {\rm Tr} \left\{ P \left[\partial_s P, {\cal D}_5 P\right]  
+ (\partial_s A_5) \, P \right\}
\, a_5 \, \Delta s ,  
\end{equation}
and if we replace $\partial_s$ to the covariant derivative ${\cal D}_s$,
it exactly coincides with
the $4+2$-dimensional topological 
field \cite{nonabelian-chiral-gauge-theory}. 

Finally, we note that the gauge anamaly 
obtained from the asymptotic value of the Chern-Simons current
in the domain wall fermion 
is related to the gauge anomaly expressed by the overlap 
Dirac operator $D$ \cite{kikukawa-noguchi}:
\begin{eqnarray}
\label{eq:CS-current-gauge-anomaly-1}
\lim_{N\rightarrow \infty}  J_5^a(x,N) &=& 
\left. 
-{\rm Tr}\left\{ T^a \gamma_5 D \right\}(x,x)
\right\vert_{U_k^1} , \\
\label{eq:CS-current-gauge-anomaly-2}
\lim_{N\rightarrow \infty}  J_5^a(x,-N+1) &=& 
\left. 
-{\rm Tr}\left\{ T^a \gamma_5 D \right\}(x,x)
\right\vert_{U_k^0} .
\end{eqnarray}
To see this, we infer from the locality property 
of the Chern-Simons current that 
the fifth component at $t=N$ 
\begin{eqnarray}
J_5^a(x,N) &=& 
{\rm Tr} 
\left\{ 
T^a\frac{1}{2}(\gamma_5-1) 
\left.
\left( \frac{1}{D_{\rm 5w} -m_0} \right) 
\right\vert_{AP}(z+\hat 5,z)
\right.
\nonumber\\
&& \quad +
\left.
\left.
T^a\frac{1}{2}(\gamma_5+1) 
\left.
\left(\frac{1}{D_{\rm 5w} -m_0} \right)
 \right\vert_{AP}(z,z+\hat 5)
\right\}
\right\vert_{t=N} \nonumber\\
\end{eqnarray}
becomes independent of the interpolation 
in the limit $N \rightarrow \infty$, depending 
only on the asymptotic value of the gauge field at $t=N$, $U_k^1(x)$. 
Then using the formula of the overlap Dirac operator 
expressed in terms of the inverse five-dimensional 
Wilson-Dirac 
operator \cite{kikukawa-noguchi,locality-in-dwf} \footnote{In 
\cite{locality-in-dwf}, domain wall fermion is 
defined vector-likely in the interval $t \in [-N+1,N]$. 
The interval should be extended to $t \in [-N+1,3N]$ in our case.}
given by
\begin{eqnarray}
\label{eq:effective-Dirac-operator-inverse-5D}
D
&=& 
\lim_{N\rightarrow \infty}
\left\{ 
1- 
P_R 
\left.
\left( \frac{1}{D_{\rm 5w} -m_0} \right) 
\right\vert_{AP}(N,N)
%\left( \overline{D}_{\rm 5w}- m_0\right) ^{-1}
P_L 
\right.
\nonumber\\
&& \qquad \qquad
- P_L 
\left.
\left( \frac{1}{D_{\rm 5w} -m_0} \right) 
\right\vert_{AP}(-N+1,-N+1)
%\left(\overline{D}_{\rm 5w}- m_0 \right)^{-1}_{-T+1,-T+1} 
P_R 
\nonumber\\
&& \qquad \qquad
- P_R 
\left.
\left( \frac{1}{D_{\rm 5w} -m_0} \right) 
\right\vert_{AP}(N,-N+1)
%\left(\overline{D}_{\rm 5w}- m_0\right)^{-1}_{T,-T+1} 
P_R
\nonumber\\
&& \qquad \qquad
\left.
- P_L 
\left.
\left( \frac{1}{D_{\rm 5w} -m_0} \right) 
\right\vert_{AP}(-N+1,N)
%\left(\overline{D}_{\rm 5w}- m_0\right)^{-1}_{-T+1,T} 
P_L 
\right\},
\nonumber\\
\end{eqnarray}
we obtain Eqs.~(\ref{eq:CS-current-gauge-anomaly-1}) and 
(\ref{eq:CS-current-gauge-anomaly-2}).
The gauge anomalies 
Eqs.~(\ref{eq:CS-current-gauge-anomaly-1})
and (\ref{eq:CS-current-gauge-anomaly-2})  can be evaluated in the
classical continuum limit as 
in \cite{kikukawa-yamada,aoyama-kikukawa} 
(see also \cite{adams,suzuki,fujikawa,chiu})
and the earlier calculation \cite{5dim-CS-in-domain-wall-fermion} 
is reproduced.

\section{Conclusion}
\label{sec:conclusion}
\reseteqnum

The introduction of chirally asymmetric gauge-couplings
to the chiral zero modes of domain wall fermion,
as the original proposal by Kaplan, 
inevitably makes the system five-dimensional. 
We have shown that the five-dimensional dependence
can be compensated by the local and gauge-invariant 
counter term in anomaly-free chiral gauge theories.

The chiral structure of the dimensionally reduced low energy 
effective action of the chiral zero modes can be understood 
again by the Ginsparg-Wilson relation. In fact, it
provides a concrete example of the gauge-invariant 
construction of the chiral gauge theories based on the
Ginsparg-Wilson relation, where the continuous 
interpolation in the space of the gauge fields 
is partly replaced by the discrete step-wise interpolation. 
Hope is that such discrete treatment of the
interpolation of the gauge fields
would be useful for a practical implementation 
of the gauge-invariant lattice chiral gauge theories.  

We note that in the gauge-invariant construction with the 
domain wall fermion, 
the Ginsparg-Wilson relation is not used explicitly, 
in sharp contrast to the invariant construction by L\"uscher
based on the Ginsparg-Wilson relation.
This is because 
one of the points of the invariant construction is 
the formulation of the integrability condition 
in the space of gauge fields and the idea 
behind it is generic. The local cohomology problem follows 
from the integrability condition, as long as 
the requirement of locality is fulfilled.

In this respect, we also note that
our construction is applicable to
any domain wall fermion theory defined with a proper 
five-dimensional Dirac operator of the structure
\begin{equation}
D_{\rm 5} (- m_0)=
\left( D_{\rm 4}(- m_0) +1 \right) \delta_{ts} 
- P_L %U_5(t)^{-1} 
\delta_{t+1,s} 
- P_R \delta_{t,s+1}, %U_5(s)
\end{equation}
\begin{equation}
D_{\rm 4}(- m_0)^\dagger = \gamma_5   D_{\rm 4}(- m_0) \gamma_5.
\end{equation}
Such a five-dimensional fermion theory 
can lead to a certain four-dimensional lattice Dirac
operator satisfying the Ginsparg-Wilson relation 
\cite{dwf-practicality,locality-in-dwf,borici}\footnote{
Bori\c ci's five-dimensional implementation of the
overlap Dirac operator with the Hermitian Wilson-Dirac 
operator ($a_5=0$) has this structure.}.
In this sense, our construction partly
shares the general applicability with the gauge-invariant construction
based on the Ginsparg-Wilson 
relation.\footnote{The complex phase of the determinant of 
such a generic
five-dimensional Dirac operator (with a negaive mass) can also 
produce the Chern-Simons term. 
This class of  
lattice Chern-Simons terms would be understood in relation to 
the Ginsparg-Wilson relation in five-dimensions
recently discussed 
by Bietenholtz and Nishimura \cite{GW-odd-dim},
since
it is straightforward to 
construct the five-dimensional overlap Dirac 
operator \cite{overlap-D-odd-dim} 
from such a generic five-dimensional Dirac operator.
}

In the original proposal by Kaplan \cite{domain-wall-fermion}, the
dynamical  treatment of the five-dimensional gauge field was also
intended. The question of this ambitious attempt is still open.

%Nielsen-Foerster mechanism, how bad ?

\section*{Acknowledgments}

The author would like to thank M.~L\"uscher and H.~Suzuki
for valuable discussions and comments. 
He is also grateful to 
P.~Hern\'andes, M.~L\"uscher, 
K.~Jansen, H.~Wittig, H.~Suzuki and K.~Tobe 
for their kind helps during his stay at CERN.
Y.K. is supported in part by Grant-in-Aid 
for Scientific Research of Ministry of Education (\#10740116).

\appendix

\section*{Appendix}
\reseteqnum

\section{Transfer matrix of 5-dim. Wilson fermion}
\label{app:transfer-matrix}
\reseteqnum

The transfer matrix of the five-dimensional Wilson fermion 
is given in the chiral basis of gamma matrices as follows.
\begin{equation}
\label{eq:transfer-matrix}
T = e^{- H} 
= 
\left(
\begin{array}{cc} \frac{1}{B} & - \frac{1}{B} C \\
                 -C^\dagger \frac{1}{B} 
& B + C^\dagger \frac{1}{B} C 
\end{array} \right),
\end{equation}
where $C$ and $B$ are two by two matrices in the spinor space 
which define the four-dimensional Wilson-Dirac operator $D_{\rm w}$ as
\begin{equation}
  D_{\rm w} - m_0 +1 = 
\left(
\begin{array}{cc} B & C \\
                 -C^\dagger & B 
\end{array} \right).
\end{equation}
Explicitly, they are given as follows:
\begin{eqnarray}
\label{eq:operator-C}
  C &=& \, \sigma_\mu \, 
\frac{1}{2}\left(\nabla_\mu+\nabla_\mu^\ast\right), \\
\label{eq:operator-B}
 B &=& 1 + \left( 
-\frac{1}{2} \nabla_\mu\nabla_\mu^\ast - m_0\right).
\end{eqnarray}

For the gauge field satisfying the admissibility condition
Eq.~(\ref{eq:smooth-gauge-field-at-boundaries}), 
the Hamiltonian defined through the transfer matrix
\begin{equation}
H=-\ln T
\end{equation}
has a finite gap \cite{boundary-fermion,
locality-of-overlap-D,bound-neuberger}.

\section{Evaluation of the partition functions of 
5-dim. Wilson fermions}
\reseteqnum
\label{app:determinant-of-DWF}

In this appendix, we describe the calculation of the functional
determinant of the five-dimensional Wilson-Dirac fermion, 
in the cases with the anti-periodic boundary condition and 
Dirichlet boundary condition in the fifth dimension.
Here we follow the method given by Neuberger in \cite{truncated-overlap}, 
with a slight extension to include the fifth component of the 
five-dimensional gauge field.  
More generic method has been given by L\"uscher 
\cite{luscher-dwf,dwf-practicality}.

Let us consider the five-dimensional Wilson-Dirac operator 
\begin{equation}
W = %D_{\rm 5w} - {m_0} =  
\left( D_{\rm w} - m_0 +1 \right) \delta_{ts} 
- P_L U_5(t)^{-1} \delta_{t+1,s} 
- P_R \delta_{t,s+1} U_5(s) ,
\end{equation}
where $t,s \in [-T+1,T]$ for Dirichlet boundary condition and
$t,s \in [-T+1,3T]$ for anti-periodic boundary condition.
We denote the size of the fifth dimension as $N$ 
in both cases. ($N$ is an even integer.)

In the chiral basis of the gamma matrices, $W$ is written 
explicitly in the matrix form as follows:
\begin{eqnarray*}
W&=&
\left( 
  \begin{array}{cccc}
\left( \begin{array}{cc} B_1 & C_1 \\ - C_1^\dagger & B_1 \end{array}\right) &
\left[ \begin{array}{cc} 0 & 0 \\ 0 & -U_{5,1} \end{array}\right] & 
\cdots & 
\left[ \begin{array}{cc} +Y & 0 \\ 0 & 0 \end{array}\right] \\
\left[ \begin{array}{cc} -U_{5,1}^{-1} & 0 \\ 0 & 0 \end{array}\right] &
\left( \begin{array}{cc} B_2 & C_2 \\ - C_2^\dagger & B_2 \end{array}\right) & 
\ddots &
\vdots \\
\vdots &
\ddots &
\ddots &
\left[ \begin{array}{cc} 0 & 0 \\ 0 & -U_{5,N-1} \end{array}\right] \\
\left[ \begin{array}{cc} 0 & 0 \\ 0 & +X\end{array}\right] &
\cdots & 
\left[ \begin{array}{cc} -U_{5,N-1}^{-1} & 0 \\ 0 & 0 \end{array}\right]  &
\left( \begin{array}{cc} B_N & C_N \\ - C_N^\dagger & B_N \end{array}\right) \\
  \end{array}
\right),
\end{eqnarray*}
where
$X=0$, $Y=0$  for Dirichlet boundary condition and 
$X=U_{5,N}$, $Y=U_{5,N}^{-1}$ for anti-periodic boundary condition.
In order to make $W$ almost lower tridiagonal,
we first exchanges the right- and left-handed component columns
for each $t$. Then we move the leftmost column to the place of
the rightmost column.
\begin{eqnarray*}
W &\Rightarrow&
\left( 
  \begin{array}{cccc}
\left( \begin{array}{cc} C_1 & B_1 \\ B_1 & - C_1^\dagger \end{array}\right) &
\left[ \begin{array}{cc} 0 & 0 \\ -U_{5,1} & 0 \end{array}\right] & 
\cdots & 
\left[ \begin{array}{cc} 0 & +Y \\ 0 & 0\end{array}\right] \\
\left[ \begin{array}{cc} 0 & -U_{5,1}^{-1} \\ 0 & 0 \end{array}\right] &
\left( \begin{array}{cc} C_2 & B_2 \\ B_2 & - C_2^\dagger \end{array}\right) & 
\ddots &
\vdots \\
\vdots &
\ddots &
\ddots &
\left[ \begin{array}{cc} 0 & 0 \\ -U_{5,N-1} & 0 \end{array}\right] \\
\left[ \begin{array}{cc} 0 & 0 \\ +X & 0 \end{array}\right] &
\cdots & 
\left[ \begin{array}{cc} 0 & -U_{5,N-1}^{-1} \\ 0 & 0 \end{array}\right]  &
\left( \begin{array}{cc} C_N & B_N \\ B_N & - C_N^\dagger \end{array}\right) \\
  \end{array}
\right)
\\
&&\\
&&\\
&\Rightarrow&
\left( 
  \begin{array}{cccc}
\left( \begin{array}{cc} B_1 & 0 \\ - C_1^\dagger & -U_{5,1} \end{array}\right) &
\cdots & \cdots & 
\left[ \begin{array}{cc} 
+Y& C_1 \\ 0 & B_1 \end{array}\right] \\
\left[ \begin{array}{cc} -U_{5,1}^{-1} & C_2 \\ 0 & B_2 \end{array}\right] &
\left( \begin{array}{cc} B_2 & 0 \\ - C_2^\dagger & -U_{5,2} \end{array}\right) & 
\cdots &
\vdots \\
\vdots &
\left[ \begin{array}{cc} -U_{5,2}^{-1} & C_3 \\ 0 & B_3 \end{array}\right] &
\ddots &
\vdots \\
\vdots &
\cdots &
\ddots &
\left( \begin{array}{cc} B_N & 0 \\ - C_N^\dagger & 
+X \end{array}\right) 
  \end{array}
\right)
\end{eqnarray*}
We then introduces the following abbreviations for the blocked matrix
elements.
\begin{eqnarray}
&&\alpha_t \equiv
\left( 
\begin{array}{cc} B_t & 0 \\ - C_t^\dagger & -U_{5,t} \end{array}
\right)
\quad
\alpha_X \equiv
\left( 
\begin{array}{cc} B_N & 0 \\ - C_N^\dagger & { X }\end{array}
\right)
\qquad (X=0 \, ({\rm Dir.}), \ U_{5,N} \, ({\rm AP}))
\nonumber\\
&&
\beta_t \equiv
\left[ 
\begin{array}{cc} -U_{5,t-1}^{-1} & C_t \\ 0 & B_t \end{array}
\right]
\qquad
\beta_Y \equiv
\left[ 
\begin{array}{cc} { Y} & C_1 \\ 0 & B_1 \end{array}
\right] 
\quad \qquad (Y=0 \, ({\rm Dir.}), \ U_{5,N}^{-1}\, ({\rm AP}))
\nonumber\\
\end{eqnarray}
Using these, $W$ assumes the following form,
\begin{eqnarray*}
W&\Rightarrow& 
\left( 
  \begin{array}{cccc}
\alpha_1 &
\cdot & \cdot & \beta_Y \\
\beta_2 &
\alpha_2 &
\cdot &
\cdot \\
\cdot&
\ddots &
\ddots &
\cdot \\
\cdot &
\cdot &
\ddots &
\alpha_N \\
  \end{array}
\right) 
\end{eqnarray*}

In order to take account of the boundary element, $\beta_Y$, 
we assume the following factorization 
\begin{eqnarray*}
%&\Rightarrow& 
\left( 
  \begin{array}{cccc}
\alpha_1 &
\cdot & \cdot & \beta_Y \\
\beta_2 &
\alpha_2 &
\cdot &
\cdot \\
\cdot&
\ddots &
\ddots &
\cdot \\
\cdot &
\cdot &
\ddots &
\alpha_X \\
  \end{array}
\right) 
=
\left( 
  \begin{array}{cccc}
\alpha_1 &
\cdot & \cdot & 
\cdot \\
\beta_2 &
\alpha_2 &
\cdot &
\cdot \\
\cdot &
\ddots &
\ddots &
\cdot \\
\cdot &
\cdot &
\ddots &
\alpha_X \\
  \end{array}
\right)
\times
\left( 
  \begin{array}{cccc}
1 &
\cdot & \cdot & 
-V_1 \\
\cdot &
1 &
\cdot &
-V_2 \\
\cdot &
\cdot &
\ddots &
\vdots \\
\cdot &
\cdot &
\cdot &
1-V_N \\
  \end{array}
\right),
\end{eqnarray*}
and consider the
recursion equations for the elements $V_t$:
\begin{eqnarray*}
-\alpha_1 V_1 &=& \beta_Y , \\
-\beta_2 V_1 -\alpha_2 V_2 &=& 0, \\
\vdots \\
%-\beta_3 V_2 -\alpha_3 V_3 &=& 0 \\
-\beta_{N-1} V_{N-2} -\alpha_{N-1} V_{N-1} &=& 0 ,\\
-\beta_N V_{N-1} +\alpha_X (1-V_N) &=& \alpha_X .
\end{eqnarray*}
These equations can easily be solved to get
\begin{equation}
V_N = \alpha_X^{-1} \alpha_N \cdot \prod_{t=1}^N
\left\{ -\alpha_t^{-1} \beta_t\right\} \cdot \beta_1^{-1} \beta_Y   .
\end{equation}
Then the determinant of $W$ is evaluated as follows:
\begin{eqnarray*}
\det \left(D_{\rm 5w} - {m_0} \right)_{X,Y}
&=& %(-1)^{q(N+1)}
\prod_{t=1}^{N-1} \det \alpha_t \cdot \det \alpha_X \det (1-V_N)
\\
&=& %(-1)^{q(N+1)}
\prod_{t=1}^{N} \det \alpha_t \cdot 
\det (\alpha_N^{-1}\alpha_X-\alpha_N^{-1}\alpha_X V_N).
\end{eqnarray*}
Here we have omitted the sign factor given by $(-1)^{q(N+1)}$ 
where $q= 2 N_c L^4$ and $N_c$ is the dimension of 
the gauge group representation, because it turns out to be unity.

The products $-\alpha_t^{-1} \beta_t$, 
$\alpha^{-1}_N \alpha_X$ and $\beta_1^{-1} \beta_Y$ 
are evaluated as 
\begin{eqnarray*}
-\alpha_t^{-1} \beta_t 
&=& 
\left(
\begin{array}{cc} 1 & 0 \\
                  0 & U_{5,t}^{-1}
\end{array} \right)
\left(
\begin{array}{cc} \frac{1}{B_t} & - \frac{1}{B_t} C_t \\
                 -C_t^\dagger \frac{1}{B_t} 
& B_t + C_t^\dagger \frac{1}{B_t} C_t 
\end{array} \right) 
\left(\begin{array}{cc} U_{5,t-1}^{-1} & 0 \\
                  0 & 1
\end{array} \right)
\\
&=& 
\left(
\begin{array}{cc} 1 & 0 \\
                  0 & U_{5,t}^{-1}
\end{array} \right)
\, T_t \, 
\left(\begin{array}{cc} U_{5,t-1}^{-1} & 0 \\
                  0 & 1
\end{array} \right),
\\
\alpha^{-1}_N \alpha_X
&=&
\left(
\begin{array}{cc} 1 & 0 \\ 0 & -U_{5,N}^{-1} X 
\end{array} \right),
\\
\beta_1^{-1} \beta_Y
&=&
\left(
\begin{array}{cc} -U_{5,0} Y & 0 \\ 0 & 1
\end{array} \right).
\end{eqnarray*}
Collecting these results, we finally obtain
\begin{eqnarray*}
\left. \det \left(D_{\rm 5w} - {m_0} \right) \right\vert_{\rm Dir.}
&=& 
\det \left( P_R +  P_L T_N \prod_{t=1}^{N-1} 
\left\{ U_{5,t}^{-1} T_t \right\} \right) 
\nonumber\\
&& \times 
\det\left(P_R + P_L \prod_{t}U_{5,t} \right)
\cdot \prod_{t=1}^N \det ( P_L + P_R B_t ) ,
\\
&&
\\
\left.
\det \left(D_{\rm 5w} - {m_0} \right)\right\vert_{\rm AP}
&=& 
\det \left( 1 +  \prod_{t=1}^N \left\{ U_{5,t}^{-1} T_t \right\} \right) 
\nonumber\\
&& 
\times 
\det\left(P_R + P_L \prod_{t}U_{5,t} \right)
\cdot \prod_{t=1}^N \det ( P_L + P_R B_t ) .
\end{eqnarray*}

\section{The partition functions in the 
limit $N\rightarrow \infty$}
\reseteqnum
\label{app:partition-function-limit}

In this appendix, we evaluate 
the partition function of the five-dimensional Wilson-Dirac 
fermions in the limit $N \rightarrow \infty$ 
and derive Eqs.~(\ref{eq:partition-function-dwf-limit}) 
and (\ref{eq:partition-function-5d-Wilson-ap-limit}).
Here we describe the case with the anti-periodic boundary 
in some detail. The case with the Dirichlet boundary condition
can be evaluated in the same manner.

As shown in the appendix~\ref{app:determinant-of-DWF},
the partition function of the five-dimensional 
Wilson-Dirac fermion with the anti-periodic boundary 
condition is given by 
\begin{eqnarray}
\label{eq:det-5dim-Wilson-fermion}
&& \left.  \det \left(D_{\rm 5w}-m_0 \right) \right\vert_{\rm AP}
/ \prod_{t=-N+1}^{3N} N_t 
\nonumber\\
&& \qquad \qquad
=
\det 
\left( 1 + \, 
          \biggl\{ \prod_{-c_2}%_{t=-\Delta+1}^{\Delta} 
                   T_t \biggr\} \, \, 
          T_1^{2(N-\Delta)} 
          \biggl\{ \prod_{c_1}%_{t=-\Delta+1}^{\Delta} 
                   T_t \biggr\} \, \, 
          T_0^{2(N-\Delta)}
\right).   \nonumber\\
\end{eqnarray}
Divided by $\det(1+T_0^{2(N-\Delta)})$, the r.h.s. can 
be rewritten as follows:
\begin{eqnarray}
\label{eq:det-5dim-Wilson-fermion-factored}
&& \det 
\left( 1 + \, 
          \biggl\{ \prod_{-c_2}%_{t=-\Delta+1}^{\Delta} 
                   T_t \biggr\} \, \, 
          T_1^{2(N-\Delta)} 
          \biggl\{ \prod_{c_1}%_{t=-\Delta+1}^{\Delta} 
                   T_t \biggr\} \, \, 
          T_0^{2(N-\Delta)}
\right) / \det\left(1+T_0^{2(N-\Delta)}\right) \nonumber\\
&& = 
\det 
\left( \frac{1}{1+T_0^{2(N-\Delta)}} + \, 
          \biggl\{ \prod_{-c_2}%_{t=-\Delta+1}^{\Delta} 
                   T_t \biggr\} \, \, 
          T_1^{2(N-\Delta)} 
          \biggl\{ \prod_{c_1}%_{t=-\Delta+1}^{\Delta} 
                   T_t \biggr\} \, \, 
         \frac{ T_0^{2(N-\Delta)}}{1+T_0^{2(N-\Delta)}}
\right)  . \nonumber\\
\end{eqnarray}
In the limit $N \rightarrow \infty$, 
the dominant term in $T_0^{2(N-\Delta)}$ is evaluated as 
\begin{eqnarray}
T_0^{2(N-\Delta)} &=& P_0 T_0^{2(N-\Delta)}
                   +(1-P_0)T_0^{2(N-\Delta)}
\nonumber\\
&=& 
 \sum_i v_i^0 \otimes {v_i^0}^\dagger e^{2(N-\Delta) | \lambda_i | }
+ {\cal O}( e^{-2(N-\Delta) \lambda_+^0 } )
\end{eqnarray}
where $v_i^0$ are eigenfunctions of $H_0=-\ln T_0$ belonging to
the negative eigenvalues $\lambda_i^0$, while $\lambda_+^0$ is
the smallest positive eigenvalue of $H_0=-\ln T_0$.
%Similar estimation holds true for $T_1^{2(N-\Delta)}$.
Then the factors involving $T_0$ in 
Eq.~(\ref{eq:det-5dim-Wilson-fermion-factored})
reduces to the projection 
operators 
\begin{equation}
\lim_{N\rightarrow \infty} \frac{
  T_0^{2(N-\Delta)}}{1+T_0^{2(N-\Delta)}} = P_0, \quad
\lim_{N\rightarrow \infty} \frac{1}{1+T_0^{2(N-\Delta)}} = 1- P_0.
\end{equation}
where
\begin{equation}
  P_0 
= \sum_i v_i^0 \otimes {v_i^0}^\dagger
= \frac{1}{2}\left(1-\frac{H_0}{\sqrt{H^2_0}} \right).
\end{equation}
Therefore, the r.h.s. of 
Eq.~(\ref{eq:det-5dim-Wilson-fermion-factored})
reduces to the following expression:
\begin{eqnarray}
{\rm (\ref{eq:det-5dim-Wilson-fermion-factored})} &=&
\det 
\left( 1-P_0 + \, 
          \biggl\{ \prod_{-c_2}%_{t=-\Delta+1}^{\Delta} 
                   T_t \biggr\} \, \, 
          T_1^{2(N-\Delta)} 
          \biggl\{ \prod_{c_1}%_{t=-\Delta+1}^{\Delta} 
                   T_t \biggr\} \, \, 
         P_0
\right)  \\
&=&
\label{eq:det-5dim-Wilson-fermion-factored-basis}
\det \left(v_i^0, 
          \biggl\{ \prod_{-c_2}%_{t=-\Delta+1}^{\Delta} 
                   T_t \biggr\} \, \, 
          T_1^{2(N-\Delta)} 
          \biggl\{ \prod_{c_1}%_{t=-\Delta+1}^{\Delta} 
                   T_t \biggr\} v_j^0 \right) .
\end{eqnarray}
In the last expression the determinant is taken about 
the indices $i$ and $j$ of basis $\left\{ v_i^0 \right\}$.

Moreover, the term $T_1^{2(N-\Delta)}$ is also
evaluated as 
\begin{eqnarray}
T_1^{2(N-\Delta)} &=& P_1 T_1^{2(N-\Delta)}
                   +(1-P_1)T_1^{2(N-\Delta)}
\nonumber\\
&=& 
 \sum_i v_i \otimes v_i^\dagger e^{2(N-\Delta) | \lambda_i | }
+ {\cal O}( e^{-2(N-\Delta) \lambda_+^1 } )
\end{eqnarray}
where $v_i$ are eigenfunctions of $H_1=-\ln T_1$ belonging to
the negative eigenvalues $\lambda_i$, while $\lambda_+^1$ 
($\lambda_-^1$) is
the smallest positive (largest negative) 
eigenvalue of $H_1=-\ln T_1$.
Then, the dominant matrix element in 
Eq.(\ref{eq:det-5dim-Wilson-fermion-factored-basis})
is given by
\begin{eqnarray}
&&\left(v_i^0, 
          \biggl\{ \prod_{-c_2}%_{t=-\Delta+1}^{\Delta} 
                   T_t \biggr\} \, \, 
          T_1^{2(N-\Delta)} 
          \biggl\{ \prod_{c_1}%_{t=-\Delta+1}^{\Delta} 
                   T_t \biggr\} v_j^0 \right)
\nonumber\\
&& =
\sum_k    
\left(v_i^0, 
          \biggl\{ \prod_{-c_2}%_{t=-\Delta+1}^{\Delta} 
                   T_t \biggr\} \, v_k^1\right) 
e^{2(N-\Delta)|\lambda_k|}
\left(v_k^1,
          \biggl\{ \prod_{c_1}%_{t=-\Delta+1}^{\Delta} 
                   T_t \biggr\} v_j^0 \right)
+ {\cal O}\left( e^{-2(N-\Delta) \lambda_+ } \right) .
\nonumber\\
\end{eqnarray}
Accordingly, the dominant contribution of the determinant in 
Eq.(\ref{eq:det-5dim-Wilson-fermion-factored-basis})
is evaluated as
\begin{eqnarray}
&&
\det \left(v_i^0, 
          \biggl\{ \prod_{-c_2}%_{t=-\Delta+1}^{\Delta} 
                   T_t \biggr\} \, \, 
          T_1^{2(N-\Delta)} 
          \biggl\{ \prod_{c_1}%_{t=-\Delta+1}^{\Delta} 
                   T_t \biggr\} v_j^0 \right)
\nonumber\\
&& 
=\det\left(1-P_0 + P_0 \biggl\{ \prod_{-c_2} T_t
    \biggr\} P_1 \biggl\{ \prod_{c_1} T_t
    \biggr\} \right)
%\det(1-P_0+P_0 T_0)^{2(T-\Delta)}
\, \det(1-P_1+P_1 T_1)^{2(N-\Delta)} 
\nonumber\\
&& \qquad
\times \left(1+ {\cal O}(e^{-2(N-\Delta) (\lambda_+^1
+|\lambda_-^1|)}\right).
\end{eqnarray}
From this result, we immediately obtain 
Eq.~(\ref{eq:partition-function-5d-Wilson-ap-limit}).

%Since $N_t$, $\det(1-P_0+P_0 T_0)$ and
%$\det(1-P_1+P_1 T_1)$ are all real, 

\section{The inverse five-dimensional Wilson-Dirac operator}
\reseteqnum
\label{app:5d-wilson-D-inverse-diff}

In this appendix, we discuss the relation between 
the inverse five-dimensional Wilson-Dirac operator
defined with the Dirichlet boundary condition and 
that defined in the infinite extent of the fifth dimension, 
and show that the replacement of 
Eq.~(\ref{eq:diff-inverse-5dim-D}) is allowed
in the limit $N \rightarrow \infty$.

For this purpose, we first note that the Dirichlet
boundary condition can be implemented 
by including the surface term
in the infinite volume. 
Namely, if we consider the five-dimensional Dirac 
fermion defined in the infinite extent of fifth dimension,
but with the couplings between
the lattice sites $(-N, -N+1)$ and between the lattice sites 
$(N, N+a_5)$ omitted,
then 
the field in the interval $[-N+1,N]$ does not
have any coupling to those outside the region and 
it is nothing but
the field defined with the Dirichlet boundary condition 
imposed at $t=-N$ and $t=N+1$.

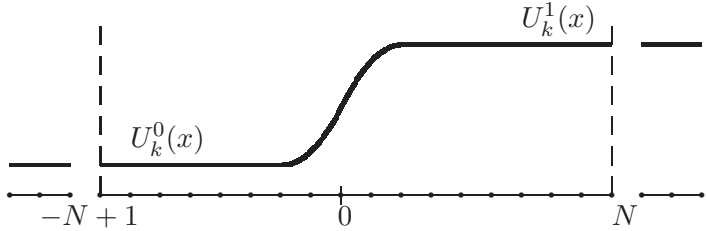
\begin{figure}[htbp]
  \begin{center}
{\unitlength 0.8mm
\begin{picture}(160,40)

\put(5,0){\line(1,0){10}}
\put(20,0){\line(1,0){85}}
\put(110,0){\line(1,0){10}}
\put(60,-1.5){\line(0,1){3}}
\put(59.5,-5){0}
%\put(120,-4){t}

%% lattice 
\multiput(5,0)(5,0){24}{\circle*{1}}
%\put(5,0){\line(0,1){30}}
\multiput(20,0)(0,5){6}{\line(0,1){3}}
%\put(120,0){\line(0,1){30}}
\multiput(105,0)(0,5){6}{\line(0,1){3}}
\put(10,-5){$-N+1$}
\put(105,-5){$N$}

% interpolation
{\linethickness{0.5mm}
\put(5,5){\line(1,0){10}}
\put(20,5){\line(1,0){30}}
\put(70,25){\line(1,0){35}} 
\put(110,25){\line(1,0){10}} 
\qbezier(50,5)(55,5)(60,15)
\qbezier(60,15)(65,25)(70,25)
}

\put(25,8){$U^0_k(x)$}
\put(90,28){$U^1_k(x)$}

\end{picture}
}
    \caption{Implementation of Dirichlet B.C. by surface interaction}
    \label{fig:dirichlet-bc-by-surface-interaction}
  \end{center}
\end{figure}

We denote the lattice Dirac operator for this fermion by
$D^\vee_{\rm 5w}$, which can be expressed with
the surface interaction as follows:
\begin{equation}
D^\vee_{\rm 5w}-m_0 =  D_{\rm 5w}-m_0 - V_{(-N+1;N)} ,
\end{equation}
where
\begin{eqnarray}
\label{eq:surface-interaction-appendix}
V_{(-N+1,N)} &=&   \left\{
- P_L \delta_{s,-N}\delta_{t,-N+1} 
- P_R \delta_{s,-N+1}\delta_{_t,-N} 
\right.
\nonumber\\
&& 
\left. \qquad \qquad
- P_L \delta_{s,N}\delta_{t,N+1} 
- P_R \delta_{s,N+1}\delta_{t,N} \right\}.
\end{eqnarray}

Then it follows immediately that 
\begin{equation}
 \frac{1}{D^\vee_{\rm 5w}-m_0} 
-\frac{1}{D_{\rm 5w}-m_0}  
=\frac{1}{D_{\rm 5w}-m_0}\, V_{(-N+1,N)} \, 
\frac{1}{D^\vee_{\rm 5w}-m_0} . 
\end{equation}
Since the inverse of $D^\vee_{\rm 5w}-m_0$ 
in the interval $[-N+1,N]$ is nothing but the 
inverse five-dimensional Wilson-Dirac operator
defined with the Dirichlet boundary condition:
\begin{equation}
  \frac{1}{D^\vee_{\rm 5w} -m_0} \,(xs,yt)
= \left. \frac{1}{ D_{\rm 5w} -m_0}\,\right\vert_{\rm Dir.}\,
(xs;yt)  \quad s,t \in [-N+1,N],
\end{equation}
we can infer that 
\begin{equation}
 \left. \frac{1}{ D_{\rm 5w} -m_0}\,\right\vert_{\rm Dir.}
-\frac{1}{D_{\rm 5w}-m_0}  
=\frac{1}{D_{\rm 5w}-m_0}\, V_{(-N+1,N)} \, 
 \left. \frac{1}{ D_{\rm 5w} -m_0}\,\right\vert_{\rm Dir.},
\end{equation}
for $s,t \in [-N+1,N]$.
Using the explicit form of the surface interaction, it can 
be rewritten further as
\begin{eqnarray}
\label{eq:diff-inverse-5dim-D-appendix}
&& \left. \frac{1}{ D_{\rm 5w} -m_0}\,\right\vert_{\rm Dir.}(xs,yt)
-\frac{1}{D_{\rm 5w}-m_0}(xs,yt)
\nonumber\\
&&=
- \frac{1}{D_{\rm 5w}-m_0}(xs;z,-N) P_L 
 \left. \frac{1}{ D_{\rm 5w} -m_0}\,\right\vert_{\rm Dir.}(z,-N+1;yt)
\nonumber\\
&& \quad
- \frac{1}{D_{\rm 5w}-m_0}(xs;z,N+1) P_R 
 \left. \frac{1}{ D_{\rm 5w} -m_0}\,\right\vert_{\rm Dir.}(z,N;yt)
\end{eqnarray}
where the summation over $z$ is understood and
$s,t \in [-N+1,N]$.

Now we consider the case where 
$s,t \in [-\Delta+1,\Delta]$. 
(See Eq.~(\ref{eq:diff-inverse-5dim-D})) .
From the exponential bound Eq.~(\ref{eq:exponetial-bound-on-5dim-D}),
the inverse five-dimensional Wilson-Dirac 
operators defined in the infinite extent of the fifth dimension
in the r.h.s. of Eq.~(\ref{eq:diff-inverse-5dim-D-appendix})
vanish identically in the limit $N\rightarrow \infty$. 
On the other hand, 
the inverse five-dimensional Wilson-Dirac 
operators defined with the Dirichlet boundary condition 
in the r.h.s. of Eq.~(\ref{eq:diff-inverse-5dim-D-appendix}) 
can be expressed in terms of the transfer matrix 
using the same technique used 
in the appendix~{\ref{app:determinant-of-DWF}} \cite{kikukawa-noguchi}.
They are given by
\begin{eqnarray}
&& 
P_L 
 \left. \frac{1}{ D_{\rm 5w} -m_0}\,\right\vert_{\rm Dir.}(z,-N+1;yt)
= +P_L \Delta(z,yt) , \\
&&P_R 
 \left. \frac{1}{ D_{\rm 5w} -m_0}\,\right\vert_{\rm Dir.}(z,N;yt)
\ \qquad = -P_R \Delta(z,yt) ,
\end{eqnarray}
where 
\begin{eqnarray}
\Delta(z,y;t)
&=&
\frac{1}{\scriptstyle P_R+
T_1^{(N-\Delta)}
\left\{
U_{5,\Delta}^{-1}\prod_{t=-\Delta+1}^{\Delta} T_tU_{5,t-1}^{-1} \right\} 
T_0^{(N-\Delta)} P_L }
\nonumber\\
&& \quad 
\times
\left\{ T_1^{(N-\Delta)}
U_{5,\Delta}^{-1}
\prod_{s=-\Delta+1}^{t} T_s U_{5,s-1}^{-1} \, \beta_t^{-1}
\right\} \qquad t \in [-\Delta+1,\Delta] .
\nonumber\\
\end{eqnarray}
For large $N$, $\Delta(z,y;t)$ can be estimated as
\begin{eqnarray}
\Delta(z,y;t)
&\simeq &
\frac{1}{\scriptstyle (1-P_1) P_R+
P_1
\left\{
U_{5,\Delta}^{-1}\prod_{t=-\Delta+1}^{\Delta} T_tU_{5,t-1}^{-1} \right\} 
T_0^{(N-\Delta)} P_L }
\nonumber\\
&& \quad 
\times
\left\{%\scriptstyle
P_1 
U_{5,\Delta}^{-1}
\prod_{s=-\Delta+1}^{t} T_s U_{5,s-1}^{-1} \, \beta_t^{-1}
\right\} %\qquad t \in [-\Delta+1,\Delta] .
\nonumber\\
&\simeq&
\frac{{\rm cofactor \ of}\left\{
(1-P_1) P_R+
P_1
\left\{
U_{5,\Delta}^{-1}\prod_{t=-\Delta+1}^{\Delta} T_tU_{5,t-1}^{-1} \right\} 
T_0^{(N-\Delta)} P_L
\right\} }{\det\left(1-P_0+P_0 T_0\right)^{(N-\Delta)}}
\nonumber\\
&&
\times
\frac{1}{
\det\left( (1-P_1) P_R+
P_1
\left\{
U_{5,\Delta}^{-1}\prod_{t=-\Delta+1}^{\Delta} T_tU_{5,t-1}^{-1} \right\} 
%T_0^{(N-\Delta)} 
P_0 P_L \right) }
\nonumber\\
&& \quad 
\times
\left\{%\scriptstyle
P_1 
U_{5,\Delta}^{-1}\prod_{s=-\Delta+1}^{t} T_tU_{5,t-1}^{-1} \beta_t^{-1}
\right\}, %\qquad t \in [-\Delta+1,\Delta] .
\end{eqnarray}
and we can infer that 
it vanishes identically in the limit $N\rightarrow
\infty$,  as long as 
\begin{equation}
\det\left( (1-P_1) P_R+
P_1
\left\{
U_{5,\Delta}^{-1}\prod_{t=-\Delta+1}^{\Delta} T_tU_{5,t-1}^{-1} \right\} 
%T_0^{(N-\Delta)} 
P_0 P_L \right) \not = 0. 
\end{equation}
Thus the r.h.s. of Eq.~(\ref{eq:diff-inverse-5dim-D-appendix}) 
vanishes in this limit. 

%%%

\end{document}